\documentclass{article}
\PassOptionsToPackage{numbers}{natbib}
\usepackage{style_file}

\usepackage[utf8]{inputenc} 
\usepackage[T1]{fontenc}    
\usepackage[hypertexnames=false]{hyperref}       
\usepackage{url}            
\usepackage{booktabs}       
\usepackage{amsmath}
\usepackage{amsfonts}       
\usepackage{nicefrac}       
\usepackage{microtype}      
\usepackage{xcolor}         
\usepackage{graphicx}
\usepackage{mathtools}
\usepackage{nameref}

\usepackage[utf8]{inputenc}
\usepackage[T1]{fontenc}

\usepackage[numbers]{natbib}
\usepackage{hyperref}

\makeatletter
\newcommand{\nociteclick}[1]{%
  \begingroup
    \renewcommand{\hyper@natlinkstart}[2]{##2}%
    \renewcommand{\hyper@natlinkend}{}%
    \cite{#1}%
  \endgroup
}
\makeatother

\makeatletter
\newcommand{\nociteclickt}[1]{%
  \begingroup
    \renewcommand{\hyper@natlinkstart}[2]{##2}%
    \renewcommand{\hyper@natlinkend}{}%
    \citet{#1}%
  \endgroup
}
\makeatother

\title{
Semantic representations emerge in biologically inspired ensembles of cross-supervising neural networks}
\author{\textbf{Roy Urbach}\\
Department of Brain Sciences\\
Weizmann Institute of Science\\
Rehovot, Israel\\
\texttt{roy.urbach@gmail.com}
\And
\textbf{Elad Schneidman} \\
Department of Brain Sciences\\
Weizmann Institute of Science\\
Rehovot, Israel\\
\texttt{elad.schneidman@weizmann.ac.il}
}

\begin{document}

\maketitle

\begin{abstract}

Brains learn to represent information from a large set of stimuli, typically by weak supervision. Unsupervised learning is therefore a natural approach for exploring the design of biological neural networks and their computations. Accordingly, redundancy reduction has been suggested as a prominent design principle of neural encoding, but its ``mechanistic'' biological implementation is unclear. Analogously, unsupervised training of artificial neural networks yields internal representations that allow for accurate stimulus classification or decoding, but typically rely on biologically-implausible implementations. We suggest that interactions between parallel subnetworks in the brain may underlie such learning: we present a model of representation learning by ensembles of neural networks, where each network learns to encode stimuli into an abstract representation space by cross-supervising interactions with other networks, for inputs they receive simultaneously or in close temporal proximity. Aiming for biological plausibility, each network has a small ``receptive field'', thus receiving a fixed part of the external input, and the networks do not share weights. We find that for different types of network architectures, and for both visual or neuronal stimuli, these cross-supervising networks learn semantic representations that are easily decodable and that decoding accuracy is comparable to supervised networks -- both at the level of single networks and the ensemble. We further show that performance is optimal for small receptive fields, and that sparse connectivity between networks is nearly as accurate as all-to-all interactions, with far fewer computations. We thus suggest a sparsely interacting collective of cross-supervising networks as an algorithmic framework for representational learning and collective computation in the brain.


\end{abstract}

\newpage 
\section{Introduction}
The brain's representations of the world, built through learning, drive our decisions and understanding. Studies of information encoding by single neurons and populations have identified canonical examples of simple representations - from low level visual feature selectivity of mammalian V1 neurons for location and orientations in visual stimuli \cite{hubel_wiesel}, to neural selectivity to shapes, concepts, and faces in ``higher'' cortical areas in the ventral stream \cite{kanwisher1997fusiform,haxby_distributed_overlapping_face_and_objects_ventral_2001,wilson_orientations_2015, downing_visbody_2001}, or neural activity in the Hippocampus that is indicative of spatial location \cite{placecells_okeefe}. Some, more complex neuronal representations (``mixed selectivity''), for example in the inferior temporal (IT) cortex, were shown to be easily decodable for stimuli identity \cite{hung2005fast}. How such representations are learned or what organizational goal led to their formation is largely unclear. 

The brain is exposed to an enormous stream of information \cite{meister_10bits}, with only weak supervision, making unsupervised learning a natural framework to consider for neural systems, and in particular, for the formation of neural representations. Yet, our understanding of unsupervised learning in the brain is limited. One unsupervised mechanism is that of Hebbian synaptic changes \cite{hebb}, in the form of spike-time-dependent plasticity \cite{ltp, stdp, synaptic_plasticity}, where synapses grow stronger (or weaker) depending on the temporal correlation between the activity of a pre-synaptic neuron and a post-synaptic one. Another mechanism is the homeostatic normalization of the incoming or outgoing synaptic weights of individual neurons \cite{turrigiano2004homeostatic,turrigiano2008self}. These mechanisms do not offer a clear functional goal for information encoding or network function -- with the exceptions of computational models of divisive normalization of neural activity \cite{heeger1996computational,carandini2012normalization} and Oja's rule \cite{oja} (a form of Hebbian learning approximation followed by normalization of the synapses that converge to the first principal component of the input).

On the functional side, the prominent ``Efficient coding'' hypothesis \cite{efficient,efficient_sounds,bialek_efficient_biometric} suggests that neurons represent stimuli in a way that reduces redundancy in information representation between neurons \cite{atick_towards_1990,atick_efficient_lgn,olshausen_sparse_1997,predictive_retina}. 
These ideas inspired methods such as blind source separation (BSS) and Independent Component Analysis (ICA) \cite{ica_french, ica_eng, fastica, ica_overview}, as well as multilayer artificial neural networks models such as {\em Autoencoders} \cite{vae,mae} that train networks to map inputs into efficient representations by minimizing the error when reconstructing the input using a decoder (``reconstruction loss''). Temporal versions of efficient coding suggested ``Predictive coding'' by neurons and populations, namely, the encoding of predicted future stimuli or the difference between the predicted stimuli and the real one (``predictive loss'') \cite{predictive_coding, predictive_testable, cont_pred, predictive_cortex, palmer_predictive_2015}.

Importantly, learning models that rely on input predictions and representational errors \cite{predictive_cortex,predictive_friston,prednet,predictive_placecells} -- either learning an efficient coding of stimuli or predictive coding -- requires access to the inputs over time, or other forms of supervision or intricate ``credit assignment''. From a biological standpoint it seems interesting and maybe more relevant to consider 
 models that could explain how neural representations are learned in the brain without the need for detailed access to the input during error propagation to modify synapses, which implies learning could or even needs to occur in the representational space that is defined by the activity of neural populations. Inspired by the intricate interactions between many brain areas \cite{human_brain_atlas,Fan2016_brainnetome,Bullmore2009_graphtheory,Joshi2022_fmriatlas}, and in particular, the interactions between cortical columns in mammalian brains \cite{boucsein_beyond_2011}, we hypothesize that the interactions between areas or modules could underlie unsupervised learning in the brain.
 
 Notably, collective learning and self-organizing in groups of animals \cite{ants_humans, ants_physics, schooling_fish, locusts_dkhili_self-organized_2017, boids_reynolds1987flocks} or interacting agents \cite{socialtaxis,altruistic} suggest ``social'' interactions as a way for distributed systems to learn efficiently, optimize performance, and a way to represent information or hold memories. 
Similarly, unsupervised learning methods used in Artificial Neural Networks (ANNs) offer a potential way to model such learning of representations, where each network learns to ``encode'' stimuli \(x\) to some multidimensional neural activity space \(\mu(x)\) by minimizing the distance between the embeddings of ``positive examples''. For example, Siamese networks \cite{siamese, siamese_lecun, siamese_sign} are two identical networks with shared weights, that learn a mapping of stimuli by minimizing the distance between an encoder’s representation of two variations of the same input \(x\) and \(x'\) (either labeled as similar or a randomly distorted version) \(\arg\min_\mu\langle distance(\mu(x),\mu(x'))\rangle_{(x,x')}\), effectively ``attracting'' the representations towards one another. Contrastive loss terms \cite{triplet, simclr, contrastive, Yerxa2023_maxmanifoldcapacity}, in which an encoder learns a mapping according to a desired contrast between the distance of representations of positive examples and negative ones are added to avoid pathological solutions of the attraction mechanism.

Critically, these methods were implemented using mechanisms that are biologically unlikely. For example, they usually include random sampling of augmentations and maskings of the input, effectively changing neuronal ``receptive fields'' or input for every training example. Another issue is the sharing of weights between encoders. Finally, the need for negative examples limits the ability of the algorithm to be local in time. Recent work focused on learning without negative examples, or non-contrastive methods. Such methods include ``VICReg'' that adds an explicit variance maximizing loss term \cite{vicreg} to avoid collapse, or ``self-distillation'' \cite{byol,noncontrastive_theo} that rely on symmetry-breaking between the Siamese networks that still share their synaptic weights \cite{byol,dino,ibot,dinov2,ijepa,dinov3}. Building a model which utilizes these ideas and is biologically-plausible, would offer possible schemes that the brain could use for unsupervised learning.

 We therefore ask here whether neural networks that get different parts of the same input, analogous to neural circuits or sub-networks in the brain with a small receptive field \cite{receptive_field}, may supervise each other to form meaningful representations of their stimuli. Compared to efficient and predictive coding models, this learning is performed without access to the input space for the error signal, making it a different and simpler explanation for representation learning in the brain. Also, unlike most deep learning methods, we train without shared weights between encoders. We then study the learning and performance of our networks using images and neural population recordings, aiming to show how biologically inspired architecture and learning method with interacting networks can result in semantically meaningful representation of stimuli.

\newpage 

\section{Results}
\label{res_vis}

\subsection{Collective Learning by cross-supervising neural networks}
\label{method}
We consider a framework for unsupervised collective learning by multiple networks that interact and train one another to encode their stimuli. The interactions between encoding networks are a form of ``internal'' cross-supervision, as each network changes its mapping from stimuli to its output, according to the activity of the other networks, and without an external ``teaching'' signal. We name this framework: ``Cooperative Learning of Semantic Representations'', or CLoSeR, for short.

Analogous to interacting networks or modules in the brain, such as cortical columns, our Artificial Neural Networks (ANNs) that encode their stimuli and interact with one another, are each defined by their own set of parameters, and their own version of the external stimulus. Thus, unlike the common weight sharing in machine learning models, here networks learn their weights independently from one another. Moreover, when presented with a stimulus \(x\), each of the \(N\) networks receives only part of it, such that network \(i\) gets as its input \(m_i(x)\), where \(m_i\) is a fixed ``mask'' of filtered version of the stimulus. The output of each network is then given by the activity of \(D\) output (or projection) neurons, and so \(\mu_i(x_b)\) is the representation or embedding of stimulus \(b\) by network \(i\), which receives as its specific input \(m_i(x_b)\). This network-specific masking of the input is analogous to the ``receptive field'' \cite{receptive_field} of neurons or networks in the brain -- that due to their connectivity and their threshold based responses, receive only part of the external input, and are selective in their spiking response. 



\begin{figure*}
     \centering
     \includegraphics[width=\textwidth]{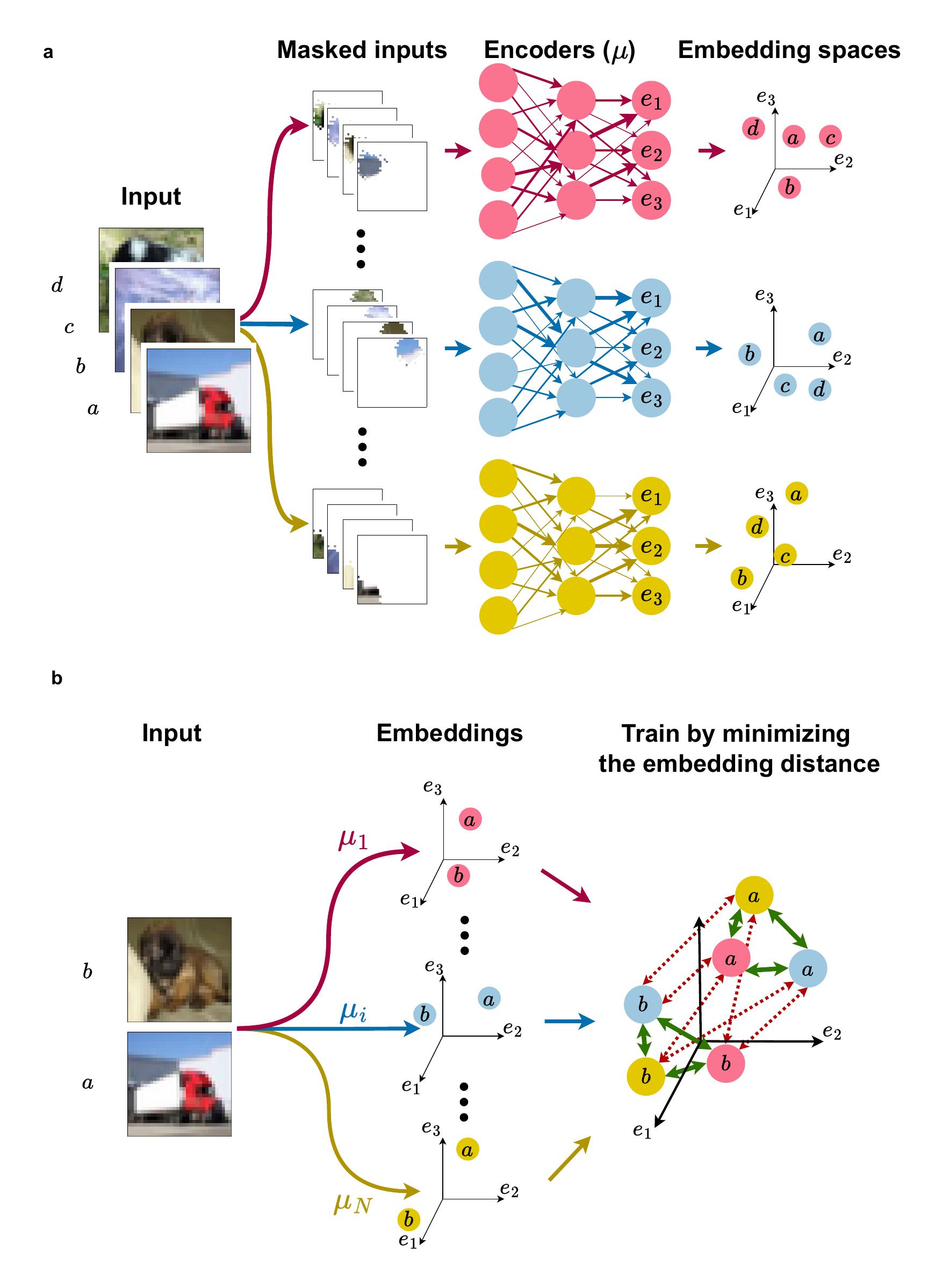}
     \caption{\textbf{Schematic illustration of the CLoSeR framework.}
     \textbf{(a)} Given an input (e.g., an image), each encoder receives a subset of the different input channels or pixels in images. The masking is randomly chosen for each network before training and remains fixed. Each encoder maps stimuli into some ``embedding space''. In this example, masking is sampled with a spatial Gaussian.
     \textbf{(b)} A schematic illustration of the training loss: For every pair of encoders \(\mu_i\) and \(\mu_j\), the encoders train to minimize the distance over their representations of the same stimulus (green arrows), and maximize distance of different ones (red arrows).}
    \label{fig_main_CLoSeR_framework}
\end{figure*}

We train the networks to find a mapping from the input space to a \(D\)-dimensional embedding space (see Fig. \ref{fig_main_CLoSeR_framework}a) that minimizes the distance between the representation of the same input sample by the different encoding networks and maximizes the distance between their responses to different stimuli (Fig. \ref{fig_main_CLoSeR_framework}b). 
Specifically, we use the Euclidean distance between the vector representations that two encoders, \(i\) and \(j\) give to two stimuli \(x_b\) and \(x_k\)
\begin{equation}\label{eq:distance}
d_{ij}(x_b,x_k)=\lVert \mu_i(x_b) - \mu_j(x_k)\rVert_2
\end{equation}
and turn it into a similarity measure:
\begin{equation}\label{eq:similarity}
    \psi_{ij}(x_b,x_k)=\exp(-\frac{d^2_{ij}(x_b,x_k)}{\tau})
\end{equation}
where \(\tau\) determines the decay of similarity with the embedding distance. 
We then use a conditional pseudo-likelihood measure (which is positive and sums to 1), calculated over the \(B\) batch examples: 
\begin{equation}\label{eq:likelihood}
    p(\mu_j(x_k) | \mu_i(x_b))=\frac{\psi_{ij}(x_b,x_k)}{\sum_{m=1}^B \psi_{ij}(x_b,x_m)}
\end{equation}

From a machine learning perspective, this can be seen as softmax of \(-\frac{d^2_{ij}(x_b,x_m)}{\tau}\) over \(m\). See more interpretations and probabilistic intuition in supplementary section: \nameref{Sup: Further intuition for the loss}).

Finally, we maximize agreement between encoders over the diagonal over the stimuli \(\{\log(p(\mu_j(x_b)|\mu_i(x_b))):b\in[B]\}\), which is akin to minimizing the loss function, 
\begin{equation}\label{eq:L_CLoSeR}
    L_{CLoSeR}=\langle-\log(p(\mu_j(x_b)|\mu_i(x_b)))\rangle_{i\neq j, b}
\end{equation}
which we perform by gradient descent on the weights of each of the networks (see Methods: \nameref{Methods: Hyperparameters and technical details}).

Intuitively, since each encoding network receives only part of the input that is defined by its unique mask \(m_i\), the networks learn an embedding that must overcome their different masking of the input and captures the underlying stimulus features that are shared between them. This is similar to common self-supervised learning algorithms \cite{siamese_face, simclr, byol, dino}, but here our masks will be biologically inspired, in the sense that they will be fixed and local. Moreover, we note that for simplicity, the architecture of all networks that we used here is similar, but it is not required by our setting.


\subsection{CLoSeR networks learn semantic representations of images}\label{res_vis_first}

We first test CLoSeR on the task of encoding static images, using the Cifar-10 dataset \cite{cifar10} of 60,000 32\(\times\)32 colored images that were curated and cropped to have an object at the center of each image that belongs to one of ten classes (e.g., dogs, trucks, birds, and ships...). For training our encoding networks, we augment the data set to avoid over-fitting \cite{alexnet}, by random flipping, rotations, and zooming-in (see Methods: \nameref{Methods: Hyperparameters and technical details}). 
Inspired by classical receptive fields, we first study the case of encoders that receive a fixed subsample of the pixels of the image, which is dictated by a 2-dimensional Gaussian with standard deviation of 2 pixels: a predefined number of pixels that participate in the mask of a network are chosen probabilistically according to their distance from the Gaussian center (see Methods: \nameref{Methods: Mask sampling}). We later show results for different masking methods. We use Multilayer Perceptrons (MLP) \cite{rosenblatt_mlp,backprop_mlp} for our encoding networks, due to their simple and biologically relevant architecture. Specifically, our encoders are composed of three hidden layers with widths \{1024, 512, 256\} and output dimension \(D=256\), with GELU activation functions \cite{gelu}. We note that the number of parameters in the first hidden layer changes according to the number of pixels the encoding network receives as input. For example, an encoding network that receives 10\% of the image has \(\sim1\) million parameters.
 
\begin{figure*}
     \centering\includegraphics[width=\textwidth]{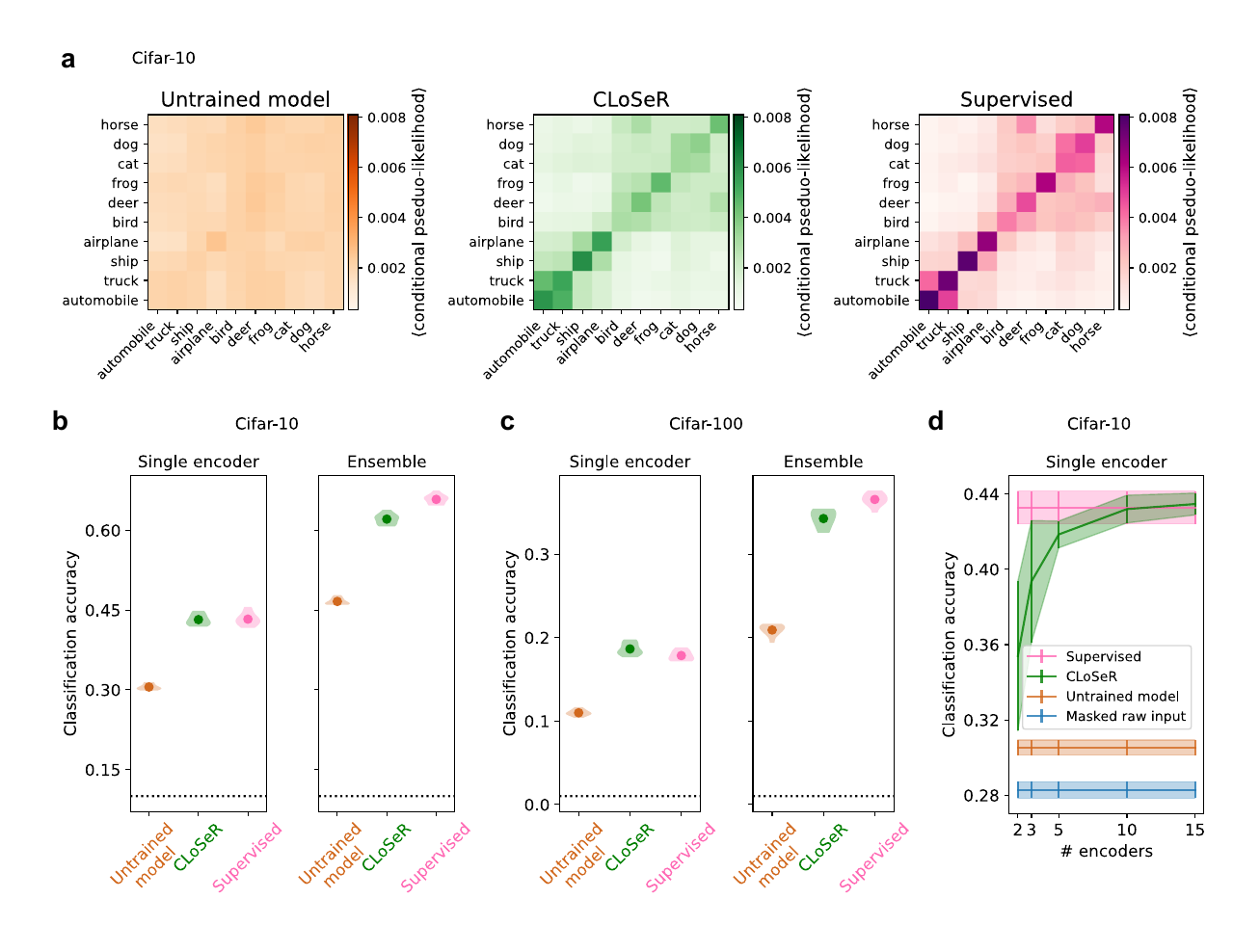}\caption{\textbf{CLoSeR with local Gaussian masking learns semantically meaningful representations.} 
     \textbf{(a)} left: The average value of the conditional pseudo-likelihood of an image from one Cifar-10 class given an image from another class, for one untrained MLP encoder that receives 10\% of the pixels of the image. Center: same for an encoder that was trained using CLoSeR. Right: Same for an encoder that was trained in a supervised manner. All panels show results for an encoder that was initialized the same way.
     \textbf{(b)} Left: The distribution over 10 sets of mean Cifar-10 accuracy values of a linear classifier decoding of each encoder's embedding for the case of CLoSer trained MLP encoders (green) untrained encoders (orange), or encoders trained by supervision (pink). The encoders in this figure receive 10\% of the pixels of the input images. Right: The distribution of accuracy values of a linear classifier decoding from the collective embeddings of the ensemble.
     The dotted black line marks chance level (10\%).
     \textbf{(c)} Same as \textbf{b} for model trained and evaluated on Cifar-100. The dotted black line marks chance level (1\%).
     \textbf{(d)} Accuracy of linear classifier decoding of a single encoding network as a function the number of encoding networks that were trained together, with 10\% of the image presented to each encoder. The 95\% CI was calculated over 10 seeds. The colors are the same as \textbf{b} and \textbf{c}, except for the blue, which is the accuracy when training a linear classifier on the masked images pixels.
     }\label{fig_main_vis_like_acc}
\end{figure*}

We trained sets of encoding networks to minimize \(L_{CLoSeR}\) for 1000 epochs, using Nadam optimizer \cite{nadam} and assessed their learned mappings by the labels of the images that they embedded close to one another. 
We measured the embedding similarity between images from pairs of image classes \(Class\:c\) and \(Class\:c'\), by the mean conditional pseudo-likelihood \(\langle p(\mu_i(x_k)|\mu_i(x_b))\rangle_{b\in Class\:c, k\in Class\:c'}\) averaged over many pairs of test images from these classes (see Methods: \nameref{Calculating class mean conditional pseudo-likelihood}). Figure \ref{fig_main_vis_like_acc}a shows an example of the similarity between images from the different classes for one encoder that was trained in a group of \(N=10\) encoders, each receiving 10\% of the patches of the image (102 pixels out of 1024). We find that the embedding of images by untrained networks is almost uncorrelated with their content, reflected by the pseudo-likelihood values they assign to their inputs, which are nearly uniformly distributed between classes. In contrast, the similarity matrix of images from different classes for an encoding network that was trained using CLoSeR has a clear diagonal structure, which means that images from the same class are embedded close to one another. The semantic organization that this network learned is reflected further in the nature of its mistakes, as it confuses, for example, automobiles and trucks more than other image categories. Interestingly, the network trained by its conspecific networks through CLoSeR, gives similar results to networks trained by supervision. 

We quantify these results over different groups of networks and masks, by the linear separability of the learned embeddings: we train a logistic regression decoder on the embedding of a single encoder to yield the class that the mapped stimulus belongs to (see details in Methods: \nameref{Methods: Training a linear classifier in the vision task}). We also trained a similar logistic regression decoder on the embedding of untrained encoders with the same architecture. The untrained encoders' representation led to decoding accuracy that is higher than chance level (\(\sim30\%\) vs. \(10\%\)), reflecting the benefits of random non-linear features (see e.g. \cite{Maoz2020_RP}). We find that the representations of every encoder that we trained using CLoSeR led to decoding accuracy that was \(\sim12\%\) higher than classifiers that we used on the embeddings of untrained encoders (fig. \ref{fig_main_vis_like_acc}b). We also trained logistic regression decoders that operate jointly on the embeddings of all the networks in the ensemble (fig. \ref{fig_main_vis_like_acc}c), and found that the decoding success of CLoSeR networks was consistent over different sampled masks -- for both the individual encoding networks and the joint set of encoders. We got similar results using CLoSeR on Cifar-100 \cite{cifar10} (see fig. \ref{fig_main_vis_like_acc}c).

We further asked how the accuracy of CLoSeR networks depend on the number of interacting encoders \(N\) that we use, and found that their performance plateaus out around 10 encoders (Fig. \ref{fig_main_vis_like_acc}d), which was the ensemble size we focused on throughout this work.

\subsection{CLoSeR with random masks is optimal for encoders with small receptive fields of different kinds}\label{res_vis_q}
\begin{figure*}
     \centering
     \includegraphics[width=\textwidth]{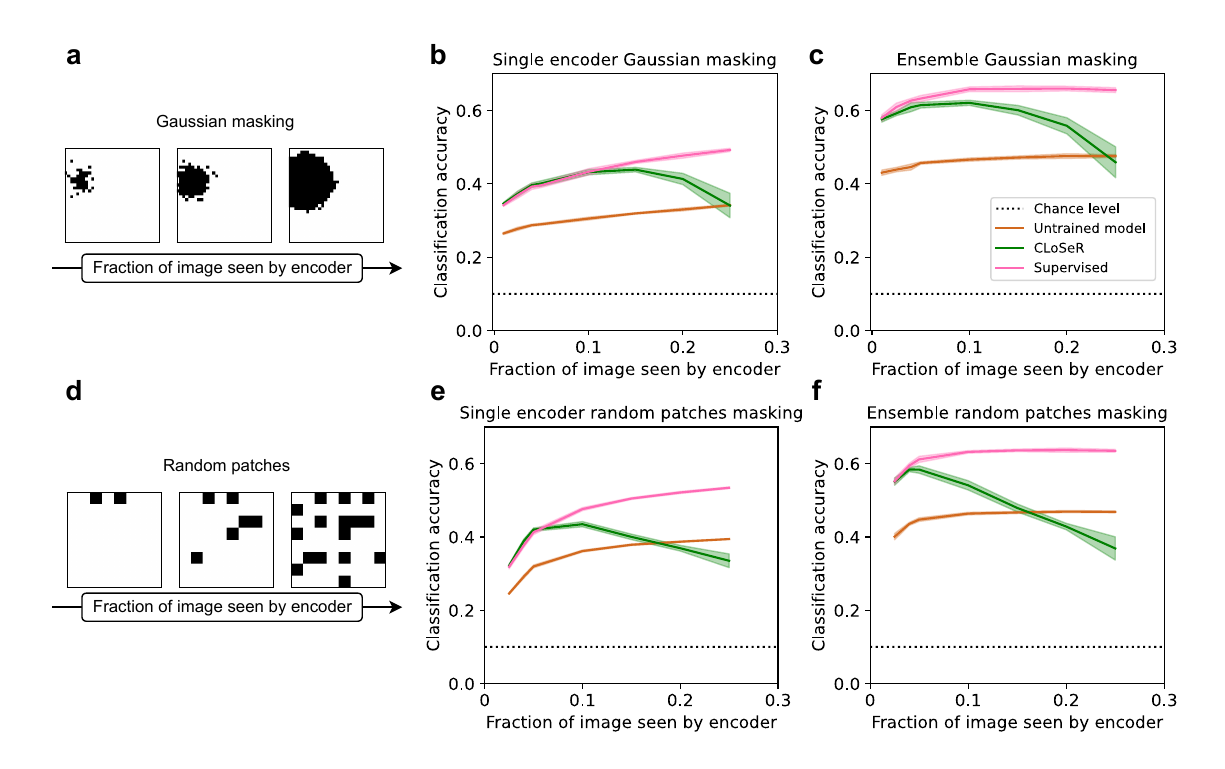}
     \caption{\textbf{Performance of CLoSeR encoders is optimal for small input masking fraction, for both Gaussian and randomly dispersed receptive fields.}
     \textbf{(a)} Example of the Gaussian masking method and how it changes as an encoder receives more pixels as input.
     \textbf{(b)} The average accuracy of a supervised linear classifier decoding the 10 MLP encoders with Gaussian masking as a function of the fraction of the image each encoder receives as input. Each line shows results for a different encoder class, and the shading shows the 95\% confidence interval over different masking seeds (\(n=10\)). 
     \textbf{(c)} Same as \textbf{b} for ensemble accuracy as a function of the fraction of image each individual encoder receives as input.
     \textbf{(d)} Example of the random patches masking method and how it changes when an encoder receives more patches as input.
     \textbf{(e-f)} Same as \textbf{b} and \textbf{c} for random patches masking.
     }
    \label{fig_main_vis_q}
\end{figure*}

Given the success of CLoSeR in training networks that give semantic representations of images, we next asked how the model's performance is affected by the size of the input masks that we used (see fig. \ref{fig_main_vis_q}a). We therefore trained different sets of encoders whose masks allow them to ``see'' different fractions of the total image, which we denote by \(q_i=\frac{|m_i|}{1024}\). For different values of \(q_i\) we trained 10 different sets of networks, each initialized using a different random seed for the masks. We find that for small enough masking fraction \(q_i\), CLoSeR leads to representations that result in higher linear classification accuracy than decoding from the mapping by untrained encoders (Fig. \ref{fig_main_vis_q}b). We further find that \(q_i \sim 15\%\) of the image was optimal for our CLoSeR setting. 
Similar to the results for the individual encoders, the ensemble embedding reaches an accuracy that is higher than that of the full image or untrained encoders (Fig. \ref{fig_main_vis_q}b). Here, however, we find an optimal value for input masking fraction around 10\%. 

The fact that there is an optimal value for \(q\) might be expected, since, due to our loss function, encoders trained by CLoSeR are likely to resort to low level features of the images as masking fraction \(q_i\) grows, rather than to learn underlying structure or semantics of the images. We also find that individual networks and the ensemble trained with CLoSeR were on par or close to networks trained by supervision for small values of the input masking fraction.

We also examined a different masking scheme, which can be considered as ``non-classical'' receptive fields for the encoders. While neurons in the visual cortex are well known to have localized selectivity for visual scenes \cite{hubel_wiesel,receptive_field}, many experiments have shown that neurons in the ventral pathway can have receptive fields with selectivity over multiple spatially discrete subregions, both in low brain areas \cite{receptive_field_multiple,allen_data} and higher areas \cite{receptive_field_mt}. We divided each image into 64 patches of size \(4\times 4\times3\) (height \(\times\) width \(\times\) color channels) \(\{x^{h,w}:(h,w)\in[8]\times[8]\}\) where \(h\) and \(w\) are the height and width indices. Each of the networks receives a subset of those patches according to its fixed input mask \(m_i(x)=\{x^{h,w}\}_{(h,w)\in [8]\times[8]}\) where the indices \((h,w)\) are sampled randomly (see fig. \ref{fig_main_vis_q}d, Methods \nameref{Methods: Mask sampling}). Thus, each network receives inputs from different parts of the image. As we have found for the case of local Gaussian masking, this random masking is optimal when the fraction of image seen by the encoders is low (see fig. \ref{fig_main_vis_q}e and f). Furthermore, in the random patches case, the optimal fraction is lower than Gaussian masking. We attribute that to the global information, which makes the task of agreement between encoders easier. We further examined how spreadingness of these random masks affects the networks' accuracy, and found that as spreadingness grows, the optimal fraction of the image that networks should see goes down (see supplementary section: \nameref{Sup: A tradeoff between receptive field size and spreadingness} and supplementary fig. \ref{fig_sup_spreadingness}). We conclude that sparse inputs to the encoders are not only more computationally efficient, but promote learning in both masking schemes.

\subsection{Sparse CLoSeR is more efficient }\label{res_vis_sparse}
In training the encoding networks above, we have used all-to-all interactions (cross-supervision) between them. Yet, in real neuronal systems, connectivity is typically sparse. We therefore explored the effect of different graphs of interactions between the encoding networks. We denote the interaction graph between them as \(G_{pull}=(V_{pull},E_{pull})\), where nodes \(V_{pull}\) represent the encoding networks, and an edge \((i,j)\in E_{pull}\) means that during training the representations of stimuli by encoder \(i\) are pulled towards those of encoder \(j\). The loss function for training the networks is then given by 

\begin{equation}\label{eq:L_CLoSeR_graph}
L_{CLoSeR}(G_{pull})=\frac{1}{|E_{pull}|} \sum_{(i,j)\in E_{pull}}\langle-\log p(\mu_i(x_b)|\mu_j(x_b))\rangle_{b}
\end{equation}
(see Methods: \nameref{Methods: CLoSeR with non-all-to-all connectivity} - for technical considerations of implementation and the difference between \(G_N^k\) full-density and all-to-all connectivity).

\begin{figure*}
     \centering
     \includegraphics[width=\textwidth]{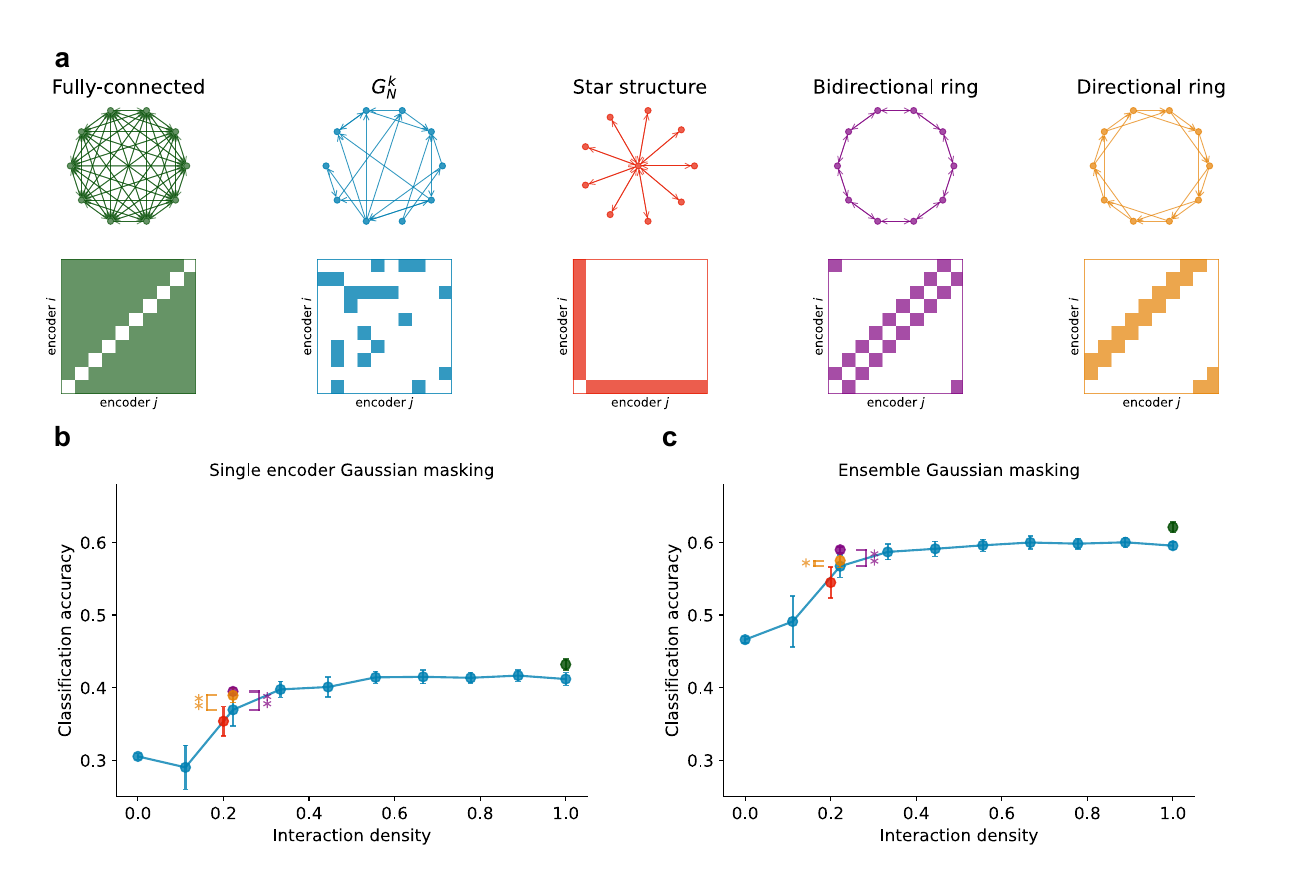}
     \caption{\textbf{CLoSeR with sparse interactions is highly efficient:}
     \textbf{(a)} Different interaction graphs we examined. Top row: illustration of the interaction graph. Bottom row: adjacency matrices of each of the interaction graphs depicted above.
     \textbf{(b)} The accuracy of single MLP Gaussian masked encoder classification as a function of the density of the interaction graph \(\nicefrac{|E|}{N\cdot(N-1)}\). 
     Different colors correspond to different graph families, as in panel a. The encoders receive 10\% of the image pixels as input. The directional ring and bidirectional ring single encoder accuracies are higher than \(G_N^k\) (paired student t-test, \(p\approx4.3\cdot10^{-3}\) and \(p\approx8.9\cdot10^{-3}\), \(df=9\)). * designates \(p<0.05\) and ** designates \(p<0.01\). 
     95\% CI is calculated over 10 seeds.
     \textbf{(c)} Same as \textbf{b} for ensemble accuracy, with \(p\approx4.4\cdot10^{-2}\) and \(p\approx4.8\cdot10^{-3}\).
     }
    \label{fig_main_sparse}
\end{figure*}

We examined four classes of interaction graphs, in addition to the fully connected one (Fig. \ref{fig_main_sparse}a): First, is the class of interaction graphs, \(G_N^k\), where \(|V|=N\) and \(|E|=k\), which are randomly sampled. The second class is a ``star'' structure, where all encoders are pulled toward a single encoder, and vice-versa: \(E_{star}=\{(i,1):i\in[2,N]\}\cup\{(1,i):i\in[2,N]\}\). The third class is that of bidirectional rings, where each encoder \(i\) is pulled towards its immediate neighbors, namely \(i-1\) and \(i+1\): \(E_{bidir}=\{(i,j):i,j\in[N]\quad\land\quad|i-j| \mod N = 1\}\). Fourth is the class of directional rings, where each encoder \(\mu_i\) is pulled towards the \(i+1\) and \(i+2\)’th encoders: \(E_{dir}=\{(i,j): i,j\in[N]\quad \land \quad 0< (j-i) \mod N \leq2\}\).

We trained the encoders using sampled \(G_N^k\) interaction graphs and different Gaussian masking seeds, and compared the accuracy of individual networks and of the ensemble for each of the different classes (Fig. \ref{fig_main_sparse}b-e). We find that the classifier accuracy of the \(G_N^k\) family begins to saturate at interaction density of \(\sim 30\%\), approaching the accuracy of the fully-connected interaction graph's classifier. Both the directional and bidirectional rings show better accuracy than the \(G_N^k\) family with the same density for individual encoders and for the ensemble.
We conclude that biologically-inspired sparse interactions between networks is sufficient to benefit from the CLoSeR framework, and that structured sparse interactions can improve performance. 

\subsection{CLoSeR learns semantic representations with other neural network architectures}
\label{res_vis_vit}
\begin{figure*}
     \centering
     \includegraphics[width=\textwidth]{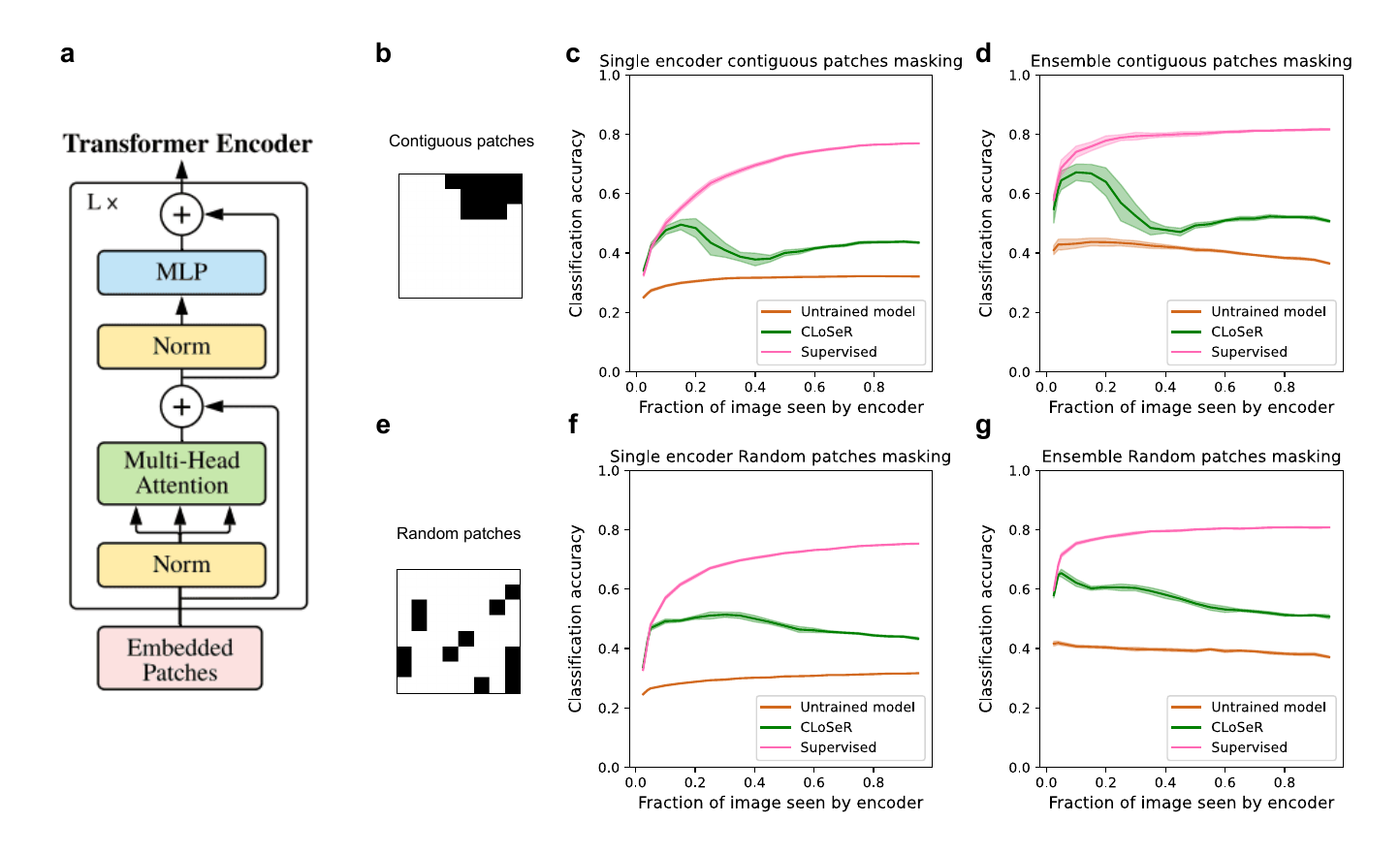}
     \caption{\textbf{Visual classification accuracy for Vision Transformer trained with CLoSeR:}
     \textbf{(a)} The Vision Transformer architecture, taken from \citet{vit}.
     \textbf{(b)} Up: example of the contiguous patches masking method, for 20\% of the image pixels.
     \textbf{(c)} Single encoder's linear classifier accuracy as a function of the image fraction seen by each ViT encoder with contiguous patches. Each line corresponds to a different training of networks. The 95\% CI was calculated over 10 seeds.
     \textbf{(d)} Same as \textbf{c} for ensemble accuracy.
     \textbf{(e)} Example of the random patches masking method for 20\% of the image pixels.
     \textbf{(f-g)} Same as \textbf{c-d}, for random patches masking.
     }
    \label{fig_main_vit}
\end{figure*}

While we have focused on MLP networks, due to their simplicity and biological plausiblity, the CLoSeR framework is applicable and beneficial for different neural architectures: We use Vision Transformers \cite{vit} for our encoding networks (see fig. \ref{fig_main_vit}a), which are the state of the art neural network architectures used for unsupervised learning of image embeddings \cite{mae, dinov2, dinov3}. The transformers regard each patch as an individual unit, and process them using an ``attention'' mechanism \cite{attn}. We picked the ViTLite-2/4 variant of Vision Transformer \cite{vitlite}, which was shown to achieve good performance on Cifar-10 in supervised and unsupervised tasks, using smaller patch-size, fewer layers, each with fewer computations, and approximately a million parameters, almost 100 time fewer parameters than \citet{vit} (see details in Methods: \nameref{Methods: Vision Transformer architecture}). Notably, Transformers are not considered biologically plausible (although there are suggestions of how glial cells, neuromodulators, and other biological processes may implement attention-like mechanisms \cite{hebattn, astroattn}), and so our focus here is on the benefits of cross-supervision learning rather than the detailed architecture.

We trained \(N=10\) ViT encoders using the CLoSeR settings and examined the learned embeddings using a linear classifier. We examined, as above, two masking methods: contiguous or random patches (see fig. \ref{fig_main_vit}b,e and Methods \nameref{Methods: Mask sampling}). Here too we find that for both masking methods, the encoders trained with CLoSeR reached better accuracy than untrained encoders, and were comparable to encoders trained in a supervised manner for small enough image fraction seen (Fig. \ref{fig_main_vit}c,d,f,g). As may be expected, the accuracy levels were slightly higher here than those of the MLP.

\subsection{Extending CLoSeR to learning of semantic representations of temporal data}\label{Learning semantic representations of temporal data}

Since the stream of natural stimuli is typically correlated over time, temporally adjacent stimuli could serve as a form of a teaching signal for the brain. Such learning could be related to synaptic plasticity rules where temporally close events induce synaptic changes \cite{hebb,stdp}, and has been suggested to be useful for learning slow features \cite{sfa}. We therefore asked  if we could adapt CLoSeR to a temporal setting and use these temporal correlations.

\begin{figure*}[h]
     \centering
     \includegraphics[width=\textwidth]{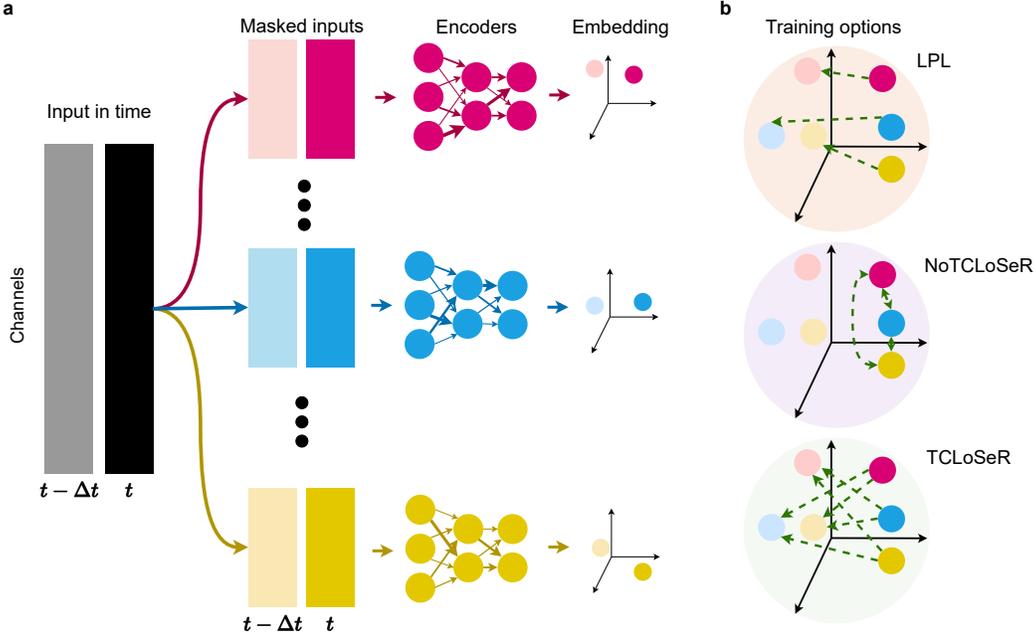}
     \caption{\textbf{A schematic illustration of a temporal variant of our objective, TCLoSeR:} 
     A CLoSeR variant in a temporal setting, where each encoder outputs an embedding per sample in time. 
     \textbf{(b)} We compare different ways to train the encoders.
     Up: LPL \cite{lpl}, where encoders train by minimizing the distance between their current and previous sample's embedding.
     Middle: NoTCLoSeR, a variant that doesn't use the temporal structure of the data, and the encoders minimize the distance between their current embedding and the other encoders' current sample's embedding.
     Bottom: TCLoSeR, the temporal variant of CLoSeR, where encoders train by minimizing the distance between their current embedding and the other encoders' previous sample's embedding (TCLoSeR), promoting both agreement between encoders and smoother embedding trajectories in time.
     }
    \label{fig_main_TCLoSeR}
\end{figure*}

We thus define a temporal variant of our algorithm, which we term TCLoSeR (see illustration in Fig. \ref{fig_main_TCLoSeR}). TCLoSeR relies on the combination of three loss terms: \(L_{tpull}\), \(L_{var}\), and \(L_{cov}\). \(L_{tpull}\) which is akin to \(L_{CLoSeR}\) (equation \ref{eq:L_CLoSeR}), is defined by the mean squared distance between encoder \(\mu_i\)'s embedding at time \(t\) and another encoder \(\mu_j\)'s embedding of the previous timestamp \(t-\Delta t\):

\begin{equation}\label{eq:L_tpull}
    L_{tpull}= \langle\lVert \mu_{i}(t)-sg(\mu_{j}(t-\Delta t)))\rVert^2_2/D\rangle_{i, j\neq i, t}
\end{equation}
where \(D\) stands for the embedding dimension, and \(sg\) stands for stopgradient. We use stopgradient so that the errors will propagate only to the current timestamp, thereby keeping this term temporally local.
The other loss terms prevent the ``collapse'' of the embeddings by the encoding networks: The first is 

\begin{equation}\label{eq:L_var}
    L_{var}= \langle-\log((\sigma^k _i)^2)\rangle_{i,k}
\end{equation}
where \((\sigma^k_{i})^2\) is the variance of the \(k\)'th output neuron of encoder \(\mu_i\) (calculated over the batch);  The second is the mean of squared covariance of two output neurons \(k\) and \(m\) from the same encoder,  
\begin{equation}\label{eq:L_cov}
    L_{cov}= \langle(\text{Cov}_{i}(k,m))^2\rangle_{i,k\neq m}
\end{equation} 
where \(i\) is the index of each of the encoders, and \(k\) and \(m\) are output neurons. 

The three loss terms are then weighted linearly:
\begin{equation}\label{eq:L_TCLoSeR}
    L_{TCLoSeR}= w_{tpull}\cdot L_{tpull} + L_{var} + 10\cdot L_{cov}
\end{equation}

where we use \(w_{tpull}=2\), since it gave the best results in preliminary trials where we explored values between \(0.1-5\). We note that the \(L_{tpull}\) term can be calculated locally in time, while the terms that mitigate the collapse of the embeddings require estimation over time. Interestingly, recent work has suggested how these collapse-avoiding terms may be approximated by Spiking Neural Networks (SNN) \cite{lpl}, implying biological plausibility.

\begin{figure*}
     \centering
     \includegraphics[width=\textwidth]{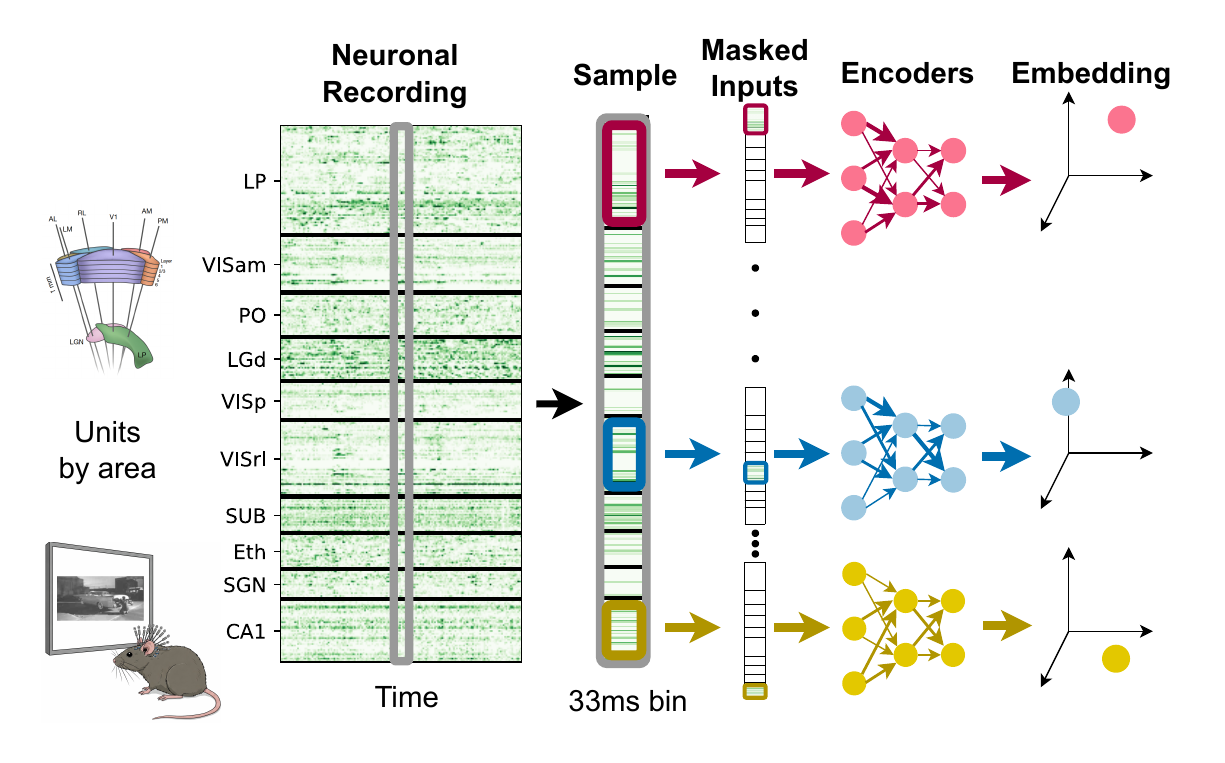}
     \caption{\textbf{An illustration of the CLoSeR setting for encoding neuronal data:}
     Given the firing rate matrix of time\(\times\)units, sorted by the area the units came from, each time bin is considered as a sample, which is masked by area (each encoder receives all units from a single brain area) and transformed by the encoder. The upper left illustration is taken from \citet{allen_data}.
     }
     \label{fig_main_neur_scheme}
\end{figure*}

We note that TCLoSeR is closely related to Latent Predictive Learning (LPL) \cite{lpl} (see fig. \ref{fig_main_neur_scheme}b up) -- a representation learning algorithm that relies on a BCM-like learning rule \cite{bcm} for the connections between neurons and stabilization terms based on temporal correlations,
that was shown to result in single encoding network learning to form semantic representation of images, with a loss function given by 
\begin{equation}\label{eq:L_LPL}
    L_{LPL}= \langle\lVert\mu_{i}(t)-sg(\mu_{i}(t-\Delta t))\rVert^2_2/D\rangle_{i,t} + L_{var} + 10\cdot L_{cov}
\end{equation}
For TCLoSeR, we used the same collapse-avoiding terms as in LPL, \(L_{var}\) and \(L_{cov}\), which in turn were adapted from \citet{vicreg}, but changed the pulling loss term (which in LPL is the mean distance between the embedding of an encoder at time \(t\) and its embedding in the previous timestamp), to the cross-encoder term \(L_{tpull}\).

To evaluate the contribution of the different terms here, we also consider an extension of the CLoSeR loss without using the temporal delay in \(L_{tpull}\), using cross-supervision at the same timestamp (see fig. \ref{fig_main_neur_scheme}b middle), and so the loss in this case is given by 

\begin{equation}\label{eq:L_NoTCLoSeR}
    L_{NoTCLoSeR}= w_{pull}\cdot L_{pull} + L_{var} + 10\cdot L_{cov}
\end{equation}

where we use, again, \(w_{pull}=2\), and 
\begin{equation}
    L_{pull} = \langle \lVert \mu_{i}(t)-sg(\mu_{j}(t)))\rVert^2_2/D\rangle_{i, j\neq i, t}
\end{equation}

\subsection{Temporal CLoSeR learns semantic representations of neural population codes}
\label{res_neur_mlp}

The success of our CLoSeR framework in teaching networks to learn representations of their stimuli, at the single network level and as a group, relied on the assumption that errors and rewards can propagate back from the output of the networks throughout the full network and back to the input layer -- as is common in many similar models  \cite{brainscore,brainscore_unsupervised,tdann}). Yet, relaxing this biologically-implausible assumption would make the framework we presented more relevant for real neural networks, and could possibly simplify the learning itself. 
Moreover, considering biologically plausible architectures or modules, raises the question of explicit representations by real neural circuits (rather than considering the external input).


We therefore evaluate TCLoSeR's performance in learning to represent neuronal activity as its input. We used the Allen Observatory’s “Visual Coding - Neuropixels” dataset \cite{allen_data}, where neural activity from hundreds of neurons was simultaneously recorded from several brain areas in mice, using Neuropixels probes, while they were watching visual stimuli. Two of the stimuli presented to the mice were scenes from the black-and-white feature movie “Touch of Evil” (1958). These movies have a ``naturalistic'' temporal structure, and similarity in time implies similarity in content, as opposed to non-natural stimuli like Gabor filters. Following previous analyses of this data set \cite{allen_data, cebra}, we assessed the accuracy of our encoding networks here in terms of their ability to classify which of the two movies did the mouse see (chance level \(\frac{2}{3}\), because one movie is twice as long as the other; see Methods: \nameref{Methods: Training a linear classifier in the neuronal task}).

\begin{figure*}
     \centering
     \includegraphics[width=\textwidth]{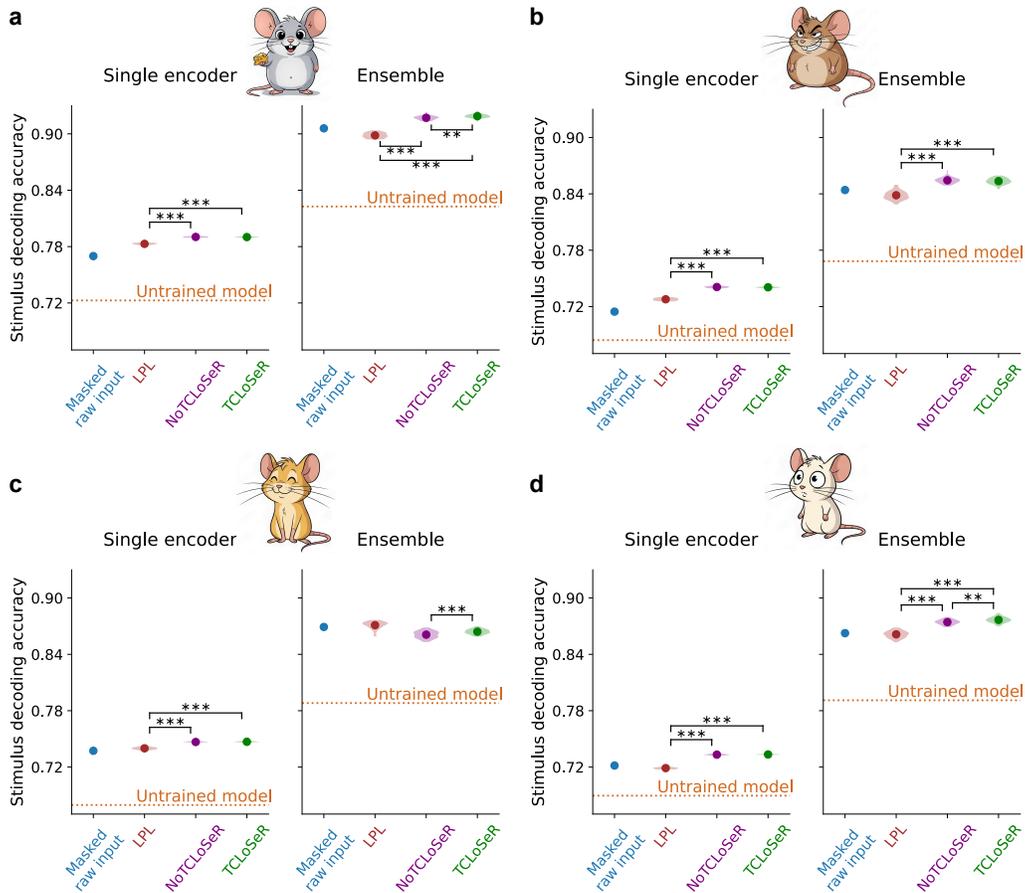}
     \caption{\textbf{Learning semantic representations of neuronal data with MLP encoders:} 
     \textbf{(a)} Left: Violinplot of the average accuracy of single MLP encoder's linear classifier is shown for different encoder types, for data from one experimental session (id 751348571). Significance calculation was done using student's t-test, and *** designates \(p<0.001\). Calculated over 30 seeds. Right: Same as left for ensemble accuracy. Chance level is \(\frac{2}{3}\).
     \textbf{(b-d)} Same as \textbf{a} for different mice session (ids 755434585, 757216464, 757970808).
     }
     \label{fig_main_neur_area}
\end{figure*}

Because of the brain's modular and anatomical structure \cite{brodmann}, we studied TCLoSeR here where each of the interacting encoding networks received as its input all units from a single brain area (Fig. \ref{fig_main_neur_area}a; see Methods: \nameref{Methods: Masking neuronal data and choosing which mice to use}). We emphasize that the number of units each encoder receives was not fixed, since each encoder received the spikes of all units from its single input area during movie trials, binned according to frame times (33ms per frame and bin). Every sample presented to the networks was of a single time bin (see figure \ref{fig_main_neur_area}a). 

To keep our algorithm biologically plausible, the encoders we used here were again MLPs, with 2 hidden layers and dimensions \{32, 32\} and output dimension 32, and a GELU activation function \cite{gelu}. We trained our \(N=10\) encoders using TCLoSeR, NoTCLoSeR, and LPL for 300 epochs, on four different mice, where we took the 10 areas with the most recorded units. We used a linear classifier using our encoders' embeddings (see Methods: \nameref{Methods: Training a linear classifier in the neuronal task}) to evaluate their performance. In three out of four mice, single encoders that were trained using temporal contiguity as the only teaching signal (LPL) achieved better accuracy than the raw neuronal activity or untrained models (see fig. \ref{fig_main_neur_area}a-d), whereas for the ensemble, LPL is better than the untrained models. Interestingly, encoders trained only with cross-supervision (NoTCLoSeR) achieved better accuracy than LPL in all mice for the single encoder, and 3 out of 4 for the ensemble. Similarly, models trained with a combination of temporal contiguity and cross-supervision (TCLoSeR) achieved better accuracy than LPL and the raw masked input in all mice for the single encoder, and three out of four mice for the ensemble. Finally, TCLoSeR achieved about the same single encoder accuracy as NoTCLoser, but significantly better ensemble accuracy in three out of four mice. We conclude that temporal contiguity and cross-supervision are not redundant in this case and that the different losses of the different models induce different solutions (see also supplementary fig. \ref{fig_sup_neur_align}a-d). Thus, the combination of temporal contiguity and interaction between encoders leads to semantic representations in a biologically-plausible architecture and setting.

\section{Discussion}
We presented a self-supervised learning algorithm in which the interactions between neural networks that encode a stimulus lead to semantically meaningful representation of their inputs by each of the individual networks as well as the collective. Unlike common self-supervised algorithms, our model adheres to biological constraints, such as non-shared weights between networks, sparse and random inputs to each encoder, and sparse connectivity between encoders. Moreover, we showed that our framework works for ``classical" local receptive fields and biologically-plausible architecture (Multilayer Perceptrons), but also for non-classical receptive fields as well as for non-biological state-of-the-art encoding networks (Transformers). Applied to naturalistic images and to temporal patterns of the activity of large populations of cortical neurons, we have found our approach to be accurate and under some conditions competitive with supervised learning. Furthermore, our algorithm shows that sparse inputs to the networks are both more efficient and promote learning, and that rather than all-to-all interactions between encoders, only a subset of interactions between networks is required to achieve near optimal performance, and structured interactions such as a bidirectional ring improved the performance of the encoders even more. 
Some of our results have corroborating evidence in previous studies. For example, in deep learning, similar sparse input masking has been shown to be effective in self-supervised tasks \cite{mae}, whereas the nature of the interaction structure between networks that is needed for efficient learning is consistent with the observed sparseness and potential design principles of neural connectomes \cite{adam_structure}. Combining its biological plausibility and performance, our approach suggests a cross-supervised self-learning algorithm for networks or modules in the brain.

One possible extension of our model is to require the embedding of network outputs to mimic the structure of early processing areas in the cortex. That way, input masking will be based on well established maps in the brain such as retinotopic maps \cite{retinotopy}, and lateral interactions can be sampled from empirical results from cortical connectomes \cite{human_connectome}. Since our model is based on the agreement between encoders, it is possible that this combination might explain phenomenon such as orientation maps in the brain \cite{orientation_maps}, which was shown to appear in similar models of learning visual representations \cite{tdann, toponet}. 

While our model had incorporated and relied on many biological features and constraints, it is not fully biologically-plausible in its current form. First, we have used feedforward deep neural networks, for simplicity and because they are much easier to train compared to other architectures. Extending this algorithm to Recurrent Neural Networks would make it more biologically relevant, and may uncover more intricate nature of the collective learning and its design in the brain. Another limitation is that our model is currently trained using a gradient descent variant with backpropagation, and the analogue of it to real neurons or even more plausible models is not straight-forward. Thus, another natural extension of our work is to consider more local and biological learning rules for relations between neurons and networks. A partial solution may come from implementing the algorithm in a layer-local and temporally-local manner. It can be extended to layer-local learning, similar to \citet{lpl}. We note that our positive examples are already local in time, while the collapse avoiding terms are still non-local. We hypothesize that using symmetry-breaking ideas from self-distillation methods will lead to a local learning rule without contrastive terms or direct variance-maximizing terms that still avoids collapsing.

More broadly, our results imply an interesting relation between the sparseness of inputs to the networks and the architecture of interactions between them. Models and analysis of collective behavior and collective learning by groups of animals \cite{ants_physics,schooling_fish} and groups of agents \cite{altruistic,socialtaxis} hints that more intricate design of interactions between networks may be useful, e.g. adding ``push" between networks (that would make them try to form distinct embeddings of the stimuli, as in \citet{altruistic}) or clustered organization of networks. Thus, exploring scaled up versions of our model, more stimulus classes, including ``pull" and ``push" mechanisms that might lead to diversification, and specialized graph topologies for the interactions between networks suggest a path towards more efficient and biological models of cross-supervision and semantic representations.

\section*{Acknowledgments}
We thank the members of the Schneidman lab for their insights and suggestions.

This work was supported by Simons Collaboration on the Global Brain grant 542997 (ES), Israel Science Foundation grant 137628 (ES), Israeli Council for Higher Education/Weizmann Data Science Research Center (ES), Martin Kushner Schnur, and Mr. \& Mrs. Lawrence Feis. ES is the incumbent of the Joseph and Bessie Feinberg Chair.

\newpage
\section*{Methods}
\label{methods}
\subsubsection*{CLoSeR with Multilayer Perceptron}\label{Methods: CLoSeR with Multilayer Perceptron}
The architecture we used for the Multilayer Perceptron (MLP) experiments in sections \ref{res_vis_first} to \ref{res_vis_sparse} on the Cifar-10 and Cifar-100 datasets is a fully-connected feed-forward MLP with three hidden layers in dimensions \{1024, 512, 256\} and output dimension 256 (same as the ViT), with GELU activation function \cite{gelu} and dropout with rate 0.1. Each encoder requires between 0.8-1.3 million parameters, depending on the fraction of images seen by the encoder.
For the neuronal task in sections \ref{res_neur_mlp}, we used MLPs with 2 hidden layers in dimensions \{32, 32\} and output dimension 32, a GELU activation function \cite{gelu} and dropout with rate 0.1. The number of parameters depends on the number of input units, and therefore ranged between 3000 to 7000.

\subsubsection*{Vision Transformer architecture}\label{Methods: Vision Transformer architecture}
Vision Transformers are a neural network architecture that adapts the Transformer model—originally developed for language tasks—to image understanding \cite{attn,vit}. 
For our image encoders in section \ref{res_vis_vit}, the architecture we used is the Vision Transformer lite model suggested by \cite{vitlite}. For computational reasons, we used the lite model with fewer attention layers than the original version (2 vs. 7), since small-scale experiments with the larger model reached similar qualitative results (albeit better performance). The internal dimension and the output dimension \(D\) we used was 256, and we set \(H=4\) attention heads and query and key dimension \(d_k=\frac{D}{H}=64\).
Since we did not want to assume prior knowledge in our network, we set the position encoding to be learnable. In total the model had approximately million parameters (whereas the original ViT-12/16 model has 85.63M). The model outputs are the accumulated activations of the \textbf{cls} token after all layers.

\subsection*{Mask sampling}\label{Methods: Mask sampling}
In this section we will explain how each masking scheme in the visual sections was implemented. Let us assume that we want to sample \(K\) pixels or \(P\) \(s\times s\) patches.

{\em Gaussian masking}: For an encoder \(\mu_i\) we first sample a center \((h_i,w_i)\) from a uniform distribution over the image pixels \(h_i\sim\mathcal{U}[H], w_i\sim\mathcal{U}[W]\), where \(H\) and \(W\) are the height and width of the images (in our case, \(32\times32\). We then define for every pixel index \((h,w)\) a probability \(p^i_{h,w}=\frac{1}{Z}\exp(-\frac{(h-h_i)^2+(w-w_i)^2}{2\sigma^2})\) where \(\sigma\) is a hyperparameter (usually we used \(\sigma=2\) except for the spreadingness experiments in supplementary section \nameref{Sup: A tradeoff between receptive field size and spreadingness}, where we also used \(\sigma=5\)), and \(Z=\sum_{(h,w)}\exp(-\frac{(h-h_i)^2+(w-w_i)^2}{2\sigma^2})\). Afterwards, we sample \(K\) pixel indices without repeats from this discrete approximation of a Gaussian.

{\em Random patches masking}: First, we divide the images into patches. The size of the patches was usually \(s=4\times4\), besides in supplementary section \nameref{Sup: A tradeoff between receptive field size and spreadingness} where we also examined patch sizes \(1\times1\) and \(2\times2\). After division we have \(\frac{H\cdot W}{s^2}\) patches. For example, in Cifar-10 we have \(32\times32\) images divided into 64 \(4\times4\) patches. Then, each encoder samples \(P\) patch indices from a uniform distribution.

{\em Contiguous patches masking}: We divide the images to patches like the random patches masking scheme. We pick \(N\) non-edge centers \(C\subset[2,\frac{H}{s}-1]\times[2,\frac{W}{s}-1]\) without repeats. We wanted the sequence of patches to be random, so for an encoder \(\mu_i\) with center \(\vec{c_i}=(h_i,w_i)\), we use the following procedure: first, define \(d^i_{h,w}=\sqrt{(h-h_i)^2+(w-w_i)^2}\) for \(h\in[0,\frac{H}{s}+1]\) and \(w\in[0,\frac{W}{s}+1]\), sample \(\epsilon^i_{h,w}\sim\mathcal{U}[0,2]\) and define \(\tilde{d}^i_{h,w}=d^i_{h,w}+\epsilon^i_{h,w}\). We then create a two-dimensional matrix of \(\tilde{d}^i_{h,w}\) according to \(h,w\), and smooth it using convolution with a \(3\times3\) Gaussian kernel, receiving a matrix \(\hat{d}^i_{h,w}\). We remove the edges to make the matrix in size \(\frac{H}{s}\times\frac{W}{s}\). Finally, we pick the indices of the top-\(P\) values of the edgeless \(\hat{d}^i_{h,w}\) matrix.

For the masking done on the neuronal data in section \ref{res_neur_mlp}, see Methods: \nameref{Methods: Masking neuronal data and choosing which mice to use}.

\subsubsection*{Calculating class mean conditional pseudo-likelihood}\label{Calculating class mean conditional pseudo-likelihood}
We measured the mean class conditional pseudo-likelihood (used in section \ref{res_vis_first}) in the following way: first, we sample \(B_{class}=50\) images per class (500 in total), and calculate \(\psi_{ii}(x_b,x_{k})\) (see equation \ref{eq:similarity}) for every encoder and pair of samples (We note that since \(\psi_{ii}(x_b,x_b)=1\), including it in our calculation, it would dampen the range of values in \(p(\mu_i(x_k)|\mu_i(x_b))\) for \(k\neq b\), and therefore we set \(\psi_{ii}(x_b,x_b)=0\) for every encoder \(i\) and image \(b\)). We calculate \(p(\mu_i(x_k)|\mu_i(x_b))\) (see equation \ref{eq:likelihood}), and then the mean class conditional pseudo-likelihood is given by
\begin{equation}
    p^i_{c'|c}(\{x_b\}_{b\in[B_{class} \cdot 10]})=\frac{1}{B_{class}\cdot(B_{class}-\delta_{c-c'})}\sum_{b=B_{class}\cdot c}^{(B_{class}+1)\cdot c -1} \sum_{k=B_{class}\cdot c'}^{(B_{class}+1)\cdot c' -1}p(\mu_i(x_k)|\mu_i(x_b))\cdot\delta_{k-b}
\end{equation}
for every pair of classes \((c,c')\in[10]\times[10]\) and \(\delta\) is Kronecker's delta. We calculate \(p^i_{c'|c}(\{x_b :b\in[B_{class} \cdot 10]\})\) for 100 random samplings of image sets, and use the average value in our analysis. Since \(p^i_{c'|c}\) relies on the similarity measure \(\psi\) from equation \ref{eq:similarity}, whose parameter \(\tau\) would affect the ``sharpness'' of \(p(\mu_i(x_k)|\mu_i(x_b))\), we used different \(\tau\) values when calculating \(\psi\) for the different models, to try and match their ``typical'' norm: We used \(\tau=10\cdot D\) for the CLoSeR models (like their training), \(\tau=0.5\cdot D\) for the untrained model and \(\tau=0.001 \cdot D\) for the supervised model. 

\subsubsection*{Training a linear classifier in the vision task}\label{Methods: Training a linear classifier in the vision task}
To assess the accuracy of each of the encoders, we trained logistic regression models using the dataset labels (Cifar-10 or Cifar-100). We use the the python package sklearn; to choose the right regularization term (\(C\) in sklearn's notation), we randomly divided the training data from the dataset into train and validation (splitting 90\% and 10\%), and then trained the logistic regression models for \(C\in\{0, 10^{-5}, 10^{-4}, 10^{-3}, 10^{-2}, 10^{-1}, 1, 10\}\). We then picked the model that gave the best accuracy on the validation data, and report its accuracy on the dataset's test examples. For the single encoder accuracy, we train the model in the following way: given images \(\{x_b\}\) and labels \(\{y_b\}\), we train the logistic regression to predict \(y_b\) from the embedding of the image by the \(i\)'th encoder \(\mu_i(x_b)\).
The ensemble's accuracy was estimated by concatenating the output of the \(N\) encoders, which gave a \(N\cdot D\) embedding for each example. On these embeddings we trained a linear classifier in the same manner as we did for single encoders above. We note that we examined different ensembling methods for the encoders' embedding, e.g., the mean or max probability prediction over encoders' linear classifiers, but these gave inferior results. For numerical stability, we trained the logistic regression on the inputs after z-score normalization (calculated over the samples).

\subsubsection*{Vision Supervised model}\label{Methods: Vision Supervised model}
For the supervised model in section \ref{res_vis_q}, we used the same architecture we used for the unsupervised case, including the masked inputs, but added an extra linear layer at the end of each encoder to yield \(o_i(x_b)\in\mathbb{R}^{10}\), from which we produce \(\tilde{o}_{ic}(x_b)=\frac{\exp(o_{ic}(x_b))}{\sum_{c'}\exp(o_{ic'}(x_b))}\) (softmax over the dimension axis), where \(\tilde{o}_{ic}(x_b)\) corresponds to the probability that the \(i\)'th encoder assigns for the \(b\)'th example to be in class \(c\). The encoders in any given set of \(N\) encoders trained independently with categorical crossentropy as the loss, using the Cifar-10 labels (or Cifar-100):
\begin{equation}
    \mathcal{L}_{supervised}(\{y_{bc}\},\{\tilde{o}_{ic}(x_b)\})=
    \langle-\sum_{c=1}^{10}y_{bc}\cdot\log(\tilde{o}_{ic}(x_b))\rangle_{b,i}
\end{equation}
where \(y_{bc}\) is a one-hot encoding of the label of the \(b\)'th example.

\subsubsection*{CLoSeR with non-all-to-all connectivity}\label{Methods: CLoSeR with non-all-to-all connectivity}
We note that \(p(\mu_j(x_b)|\mu_i(x_b))\neq p(\mu_i(x_b)|\mu_j(x_b))\) (see equation \ref{eq:likelihood}), because the denominator is not symmetric. This raises the question of whether when we optimize this term we should do it with respect to the parameters of \(\mu_i\), \(\mu_j\), or both. For the case of all-to-all interactions between encoders, we decided to optimize \(p(\mu_i(x_b)|\mu_j(x_b))\) and \(p(\mu_j(x_b)|\mu_i(x_b))\) with respect to both encoders. However, since the other interaction graphs we examined here are directional, we wanted to verify that the optimization is also directional, i.e. that the interaction \((i,j)\) only pulls \(\mu_i\) towards \(\mu_j\) and not \(\mu_j\) towards \(\mu_i\). Thus, for an interaction graph \(G=(V,E)\), we only optimize the \(i\)'th encoder according to the terms \(-\log(p(\mu_i(x_b)|\mu_j(x_b))\) where \(b\in[B]\land j\in[N]\land(i,j)\in E\) (and it is not optimized according to terms \(-\log(p(\mu_j(x_b)|\mu_i(x_b))\)). To that end, we used stopgradient in our distance calculation, which we denote by,
\begin{equation}
    \tilde{d}_{ij}(x_b,x_k)=\lVert \mu_i(x_b) - sg(\mu_j(x_k))\rVert_2
\end{equation}

The rest of calculations for \(L_{CLoSeR}(G)\) (equation \ref{eq:L_CLoSeR_graph}) remain the same except they use \(\tilde{d}\) instead of \(d\). This is why Fig. \ref{fig_main_sparse}b and Fig. \ref{fig_main_sparse}c 
 have both the blue (\(G_N^k\) with stopgrad) and the green (all-to-all without stopgrad) results for full-density.


\subsubsection*{Training a linear classifier in the neuronal task}\label{Methods: Training a linear classifier in the neuronal task}
The pipeline that was used through section \ref{res_neur_mlp} was the same as in Methods: \nameref{Methods: Training a linear classifier in the vision task}, except the labels that were the identity of the movie which was presented to the animal. The dataset was divided into train, validation, and test trials with 60\%, 20\% and 20\% division. The 30 trials were divided according to their order,
\begin{equation}
    separation(trial)=\begin{cases}
    validation & trial\mod5=1\\
    test & trial\mod5=3\\
    train & else
    \end{cases}
\end{equation}
This choice was made so we can assess generalization over different trials while keeping the trial structure, to mediate the potential issue of instability of neural responses throughout the sessions.

\subsubsection*{Masking neuronal data and choosing which mice to use}\label{Methods: Masking neuronal data and choosing which mice to use}
In sections \ref{res_neur_mlp}, when masking by area, we assume that for every unit \(u_m\) we know what brain area it was taken from - \(a_m\). We used the \(N=10\) brain areas with the largest number of units, \(A\). Then, the input to the \(i\)'th encoder is a vector of the number of spikes in the corresponding 33ms time-bin \(t\) of the units from the \(i\)'th area \(\{u_m\}\) for \(m\) s.t. \(a_m=A_i\). To use as many units as possible for a rich input, we chose the four mice that had the highest number of units in their area with the least number of units out of \(A\). The id \# of the mice sessions we used (according to the data's API) were: 751348571, 755434585, 757216464, and 757970808.

\subsubsection*{Alignment of embedding maps}\label{Methods: Alignment of embedding maps}
The alignment score measures the similarity between the embeddings of a sample by pairs of encoders, without assuming that the embedding axes are aligned: To calculate the agreement score, we sampled 50 batches \(\{B_s\}_{s\in[50]}\), each composed of \(B=128\) examples \(B_s=\{x^s_b\}\) where \(b\) belongs the \(s\)'th batch. For each batch and every pair of encoders, we computed the mean Pearson correlation between the intra-encoder distance matrices
\begin{equation}
    Alignment_{ij}(B_s)= Corr_{b,k\neq b\in[B]}(d_{ii}(x^s_b,x^s_k), d_{jj}(x^s_b,x^s_k))
\end{equation}
We then use the mean over pairs of encoders and batch samples:
\begin{equation}
    S_{align} = \frac{1}{50\cdot \binom{N}{2}}\sum_{s=1}^{50}\sum_{i\in[N]}\sum_{j\neq i}Alignment_{ij}(B_s)
\end{equation}



\subsubsection*{Hyperparameters and technical details}\label{Methods: Hyperparameters and technical details}

 {\em Encoders and training implementation}: In implementing the encoders and the training process we used tensorflow version 2.6.2.
 
 {\em Optimizer}: Throughout this work we used the optimizer Nadam \cite{nadam} with tensorflow's default parameters, which is an extension of Adam \cite{adam} with Nesterov momentum \cite{nesterov}.
 
 {\em Images augmentations}: When training our encoders on Cifar-10 or Cifar-100 in section \ref{res_vis} we used three augmentations (before masking) - random horizontal flips, random rotation (with factor 0.02 in tensorflow's implementation), random zooming (with factor 0.2 in tensorflow's implementation).
 
 {\em Values for \(\tau\) in \(L_{CLoSeR}\)}: In the definition of \(L_{CLoSeR}\) (equation \ref{eq:L_CLoSeR}), we use \(\tau=10\cdot D\) for the MLP and \(\tau=10\) for the ViT experiments.
 
 {\em Number of training epochs}: The MLP models were trained for 1000 epochs in the visual task, and ViT encoders trained for 2000 epochs for the vision task (section \ref{res_vis_vit}), and 300 epochs for the neuronal task (section \ref{res_neur_mlp}).
 
 {\em Batch size}: During the optimization process of all our models, we used batch size 128.
 
 {\em Kernel regularization}: In the neuronal task in section \ref{res_neur_mlp}, when training with TCLoSeR (equation \ref{eq:L_TCLoSeR}) and LPL (equation \ref{eq:L_LPL}), we added kernel regularization to the MLP encoders with weights \(L_1=L_2=5\cdot10^{-6}\).

 {\em Image fraction values}: For the Gaussian masking scheme, we used the following \(q_i\) values (floored according to the number of pixels): \(\{0.01, 0.025, 0.04, 0.05, 0.1, 0.15, 0.2, 0.25, 0.3\}\). Similarly, for any patch masking (both random and contiguous), we use same values (besides 0.01). For the Vision Transformers, we extended this set because the performance didn't extremely drop like MLPs and added \(\{0.35, 0.4, 0.45, 0.5, 0.55, 0.6, 0.65, 0.7, 0.75, 0.8, 0.85, 0.9, 0.95\}\).

\section*{Data availability}
{\em Cifar-10:} \url{https://www.tensorflow.org/datasets/catalog/cifar10}.

{\em Cifar-100:}  \url{https://www.tensorflow.org/datasets/catalog/cifar100}.

{\em Allen institute Visual coding Neuropixels dataset:}  \url{https://portal.brain-map.org/circuits-behavior/visual-coding-neuropixels}.

\section*{Code availability}
The code is publicly available: \url{https://github.com/roy-urbach/CLoSeR}.

\newpage
\bibliographystyle{unsrtnat}
\bibliography{bibliography}

\begin{thebibliography}{97}
\providecommand{\natexlab}[1]{#1}
\providecommand{\url}[1]{\texttt{#1}}
\expandafter\ifx\csname urlstyle\endcsname\relax
  \providecommand{\doi}[1]{doi: #1}\else
  \providecommand{\doi}{doi: \begingroup \urlstyle{rm}\Url}\fi

\bibitem[Hubel and Wiesel(1974)]{hubel_wiesel}
David~H. Hubel and Torsten~N. Wiesel.
\newblock Uniformity of monkey striate cortex: A parallel relationship between field size, scatter, and magnification factor.
\newblock \emph{Journal of Comparative Neurology}, 158, 1974.
\newblock URL \url{https://api.semanticscholar.org/CorpusID:13140630}.

\bibitem[Kanwisher et~al.(1997)Kanwisher, McDermott, and Chun]{kanwisher1997fusiform}
Nancy Kanwisher, Josh McDermott, and Marvin~M Chun.
\newblock The fusiform face area: a module in human extrastriate cortex specialized for face perception.
\newblock \emph{Journal of neuroscience}, 17\penalty0 (11):\penalty0 4302--4311, 1997.

\bibitem[Haxby et~al.(2001)Haxby, Gobbini, Furey, Ishai, Schouten, and Pietrini]{haxby_distributed_overlapping_face_and_objects_ventral_2001}
James~V. Haxby, M.~Ida Gobbini, Maura~L. Furey, Alumit Ishai, Jennifer~L. Schouten, and Pietro Pietrini.
\newblock Distributed and overlapping representations of faces and objects in ventral temporal cortex.
\newblock \emph{Science}, 293\penalty0 (5539):\penalty0 2425--2430, 2001.
\newblock \doi{10.1126/science.1063736}.
\newblock URL \url{https://www.science.org/doi/abs/10.1126/science.1063736}.

\bibitem[Wilson and Wilkinson(2015)]{wilson_orientations_2015}
Hugh~R. Wilson and Frances Wilkinson.
\newblock From orientations to objects: {Configural} processing in the ventral stream.
\newblock \emph{Journal of Vision}, 15\penalty0 (7):\penalty0 4, May 2015.
\newblock ISSN 1534-7362.
\newblock \doi{10.1167/15.7.4}.
\newblock URL \url{https://doi.org/10.1167/15.7.4}.

\bibitem[Downing et~al.(2001)Downing, Jiang, Shuman, and Kanwisher]{downing_visbody_2001}
Paul~E. Downing, Yuhong Jiang, Miles Shuman, and Nancy Kanwisher.
\newblock A cortical area selective for visual processing of the human body.
\newblock \emph{Science}, 293\penalty0 (5539):\penalty0 2470--2473, 2001.
\newblock \doi{10.1126/science.1063414}.
\newblock URL \url{https://www.science.org/doi/abs/10.1126/science.1063414}.

\bibitem[O'Keefe and Dostrovsky(1971)]{placecells_okeefe}
J.~O'Keefe and J.~Dostrovsky.
\newblock The hippocampus as a spatial map. preliminary evidence from unit activity in the freely-moving rat.
\newblock \emph{Brain Research}, 34\penalty0 (1):\penalty0 171--175, 1971.
\newblock ISSN 0006-8993.
\newblock \doi{https://doi.org/10.1016/0006-8993(71)90358-1}.
\newblock URL \url{https://www.sciencedirect.com/science/article/pii/0006899371903581}.

\bibitem[Hung et~al.(2005)Hung, Kreiman, Poggio, and DiCarlo]{hung2005fast}
Chou~P Hung, Gabriel Kreiman, Tomaso Poggio, and James~J DiCarlo.
\newblock Fast readout of object identity from macaque inferior temporal cortex.
\newblock \emph{Science}, 310\penalty0 (5749):\penalty0 863--866, 2005.

\bibitem[Zheng and Meister(2025)]{meister_10bits}
Jieyu Zheng and Markus Meister.
\newblock The unbearable slowness of being: Why do we live at 10 bits/s?
\newblock \emph{Neuron}, 113\penalty0 (2):\penalty0 192--204, 2025.
\newblock ISSN 0896-6273.
\newblock \doi{https://doi.org/10.1016/j.neuron.2024.11.008}.
\newblock URL \url{https://www.sciencedirect.com/science/article/pii/S0896627324008080}.

\bibitem[Hebb(1949)]{hebb}
Donald~O. Hebb.
\newblock \emph{The organization of behavior: {A} neuropsychological theory}.
\newblock Wiley, New York, June 1949.
\newblock ISBN 0-8058-4300-0.

\bibitem[Markram et~al.(1997)Markram, L{\"u}bke, Frotscher, and Sakmann]{ltp}
H~Markram, J~L{\"u}bke, M~Frotscher, and B~Sakmann.
\newblock Regulation of synaptic efficacy by coincidence of postsynaptic {APs} and {EPSPs}.
\newblock \emph{Science}, 275\penalty0 (5297):\penalty0 213--215, January 1997.

\bibitem[Caporale and Dan(2008)]{stdp}
Natalia Caporale and Yang Dan.
\newblock Spike timing-dependent plasticity: a hebbian learning rule.
\newblock \emph{Annu. Rev. Neurosci.}, 31\penalty0 (1):\penalty0 25--46, 2008.

\bibitem[Abbott and Nelson(2000)]{synaptic_plasticity}
L~F Abbott and S~B Nelson.
\newblock Synaptic plasticity: taming the beast.
\newblock \emph{Nat. Neurosci.}, 3 Suppl\penalty0 (S11):\penalty0 1178--1183, November 2000.

\bibitem[Turrigiano and Nelson(2004)]{turrigiano2004homeostatic}
Gina~G Turrigiano and Sacha~B Nelson.
\newblock Homeostatic plasticity in the developing nervous system.
\newblock \emph{Nature reviews neuroscience}, 5\penalty0 (2):\penalty0 97--107, 2004.

\bibitem[Turrigiano(2008)]{turrigiano2008self}
Gina~G Turrigiano.
\newblock The self-tuning neuron: synaptic scaling of excitatory synapses.
\newblock \emph{Cell}, 135\penalty0 (3):\penalty0 422--435, 2008.

\bibitem[Heeger et~al.(1996)Heeger, Simoncelli, and Movshon]{heeger1996computational}
David~J Heeger, Eero~P Simoncelli, and J~Anthony Movshon.
\newblock Computational models of cortical visual processing.
\newblock \emph{Proceedings of the National Academy of Sciences}, 93\penalty0 (2):\penalty0 623--627, 1996.

\bibitem[Carandini and Heeger(2012)]{carandini2012normalization}
Matteo Carandini and David~J Heeger.
\newblock Normalization as a canonical neural computation.
\newblock \emph{Nature reviews neuroscience}, 13\penalty0 (1):\penalty0 51--62, 2012.

\bibitem[Oja(1982)]{oja}
Erkki Oja.
\newblock Simplified neuron model as a principal component analyzer.
\newblock \emph{Journal of Mathematical Biology}, 15:\penalty0 267--273, 1982.
\newblock URL \url{https://api.semanticscholar.org/CorpusID:16577977}.

\bibitem[Barlow(1961)]{efficient}
Horace Barlow.
\newblock Possible principles underlying the transformations of sensory messages.
\newblock \emph{Sensory Communication}, 1, 01 1961.
\newblock \doi{10.7551/mitpress/9780262518420.003.0013}.

\bibitem[Lewicki(2002)]{efficient_sounds}
Michael~S Lewicki.
\newblock Efficient coding of natural sounds.
\newblock \emph{Nat. Neurosci.}, 5\penalty0 (4):\penalty0 356--363, April 2002.

\bibitem[Bialek et~al.(2007)Bialek, Steveninck, and Tishby]{bialek_efficient_biometric}
William Bialek, Rob R. de Ruyter~van Steveninck, and Naftali Tishby.
\newblock Efficient representation as a design principle for neural coding and computation, December 2007.
\newblock URL \url{http://arxiv.org/abs/0712.4381}.
\newblock arXiv:0712.4381 [q-bio].

\bibitem[Atick and Redlich(1990)]{atick_towards_1990}
Joseph~J. Atick and A.~Norman Redlich.
\newblock Towards a theory of early visual processing.
\newblock \emph{Neural Computation}, 2\penalty0 (3):\penalty0 308--320, 1990.
\newblock ISSN 1530-888X.
\newblock \doi{10.1162/neco.1990.2.3.308}.
\newblock Place: US Publisher: MIT Press.

\bibitem[Dan et~al.(1996)Dan, Atick, and Reid]{atick_efficient_lgn}
Yang Dan, Joseph~J. Atick, and R.~Clay Reid.
\newblock Efficient {Coding} of {Natural} {Scenes} in the {Lateral} {Geniculate} {Nucleus}: {Experimental} {Test} of a {Computational} {Theory}.
\newblock \emph{The Journal of Neuroscience}, 16\penalty0 (10):\penalty0 3351--3362, May 1996.
\newblock ISSN 0270-6474.
\newblock \doi{10.1523/JNEUROSCI.16-10-03351.1996}.
\newblock URL \url{https://www.ncbi.nlm.nih.gov/pmc/articles/PMC6579125/}.

\bibitem[Olshausen and Field(1997)]{olshausen_sparse_1997}
Bruno~A. Olshausen and David~J. Field.
\newblock Sparse coding with an overcomplete basis set: {A} strategy employed by {V1}?
\newblock \emph{Vision Research}, 37\penalty0 (23):\penalty0 3311--3325, December 1997.
\newblock ISSN 0042-6989.
\newblock \doi{10.1016/S0042-6989(97)00169-7}.
\newblock URL \url{https://www.sciencedirect.com/science/article/pii/S0042698997001697}.

\bibitem[Srinivasan et~al.(1982)Srinivasan, Laughlin, and Dubs]{predictive_retina}
M~V Srinivasan, S~B Laughlin, and A~Dubs.
\newblock Predictive coding: a fresh view of inhibition in the retina.
\newblock \emph{Proc. R. Soc. Lond.}, 216\penalty0 (1205):\penalty0 427--459, November 1982.

\bibitem[Herault et~al.(1985{\natexlab{a}})Herault, Jutten, and Ans]{ica_french}
Jeanny Herault, Christian Jutten, and Bernard Ans.
\newblock Detection de grandeurs primitives dans un message composite par une architeture de calcul neuromimetique en apprentissage non supervise.
\newblock \emph{10° Colloque sur le traitement du signal et des images, 1985 ; p. 1017-1022}, 01 1985{\natexlab{a}}.

\bibitem[Herault et~al.(1985{\natexlab{b}})Herault, Jutten, and Ans]{ica_eng}
Jeanny Herault, Christian Jutten, and Bernard Ans.
\newblock Detection de grandeurs primitives dans un message composite par une architeture de calcul neuromimetique en apprentissage non supervise.
\newblock \emph{10° Colloque sur le traitement du signal et des images, 1985 ; p. 1017-1022}, 01 1985{\natexlab{b}}.

\bibitem[Hyvärinen and Oja(2000)]{fastica}
A.~Hyvärinen and E.~Oja.
\newblock Independent component analysis: algorithms and applications.
\newblock \emph{Neural Networks}, 13\penalty0 (4):\penalty0 411--430, 2000.
\newblock ISSN 0893-6080.
\newblock \doi{https://doi.org/10.1016/S0893-6080(00)00026-5}.
\newblock URL \url{https://www.sciencedirect.com/science/article/pii/S0893608000000265}.

\bibitem[Naik and Kumar(2011)]{ica_overview}
G~Naik and D~Kumar.
\newblock An overview of independent component analysis and its applications.
\newblock \emph{Informatica}, 35:\penalty0 63--81, 2011.

\bibitem[Kingma and Welling(2022)]{vae}
Diederik~P Kingma and Max Welling.
\newblock Auto-encoding variational bayes, 2022.
\newblock URL \url{https://arxiv.org/abs/1312.6114}.

\bibitem[He et~al.(2021)He, Chen, Xie, Li, Dollár, and Girshick]{mae}
Kaiming He, Xinlei Chen, Saining Xie, Yanghao Li, Piotr Dollár, and Ross Girshick.
\newblock Masked {Autoencoders} {Are} {Scalable} {Vision} {Learners}, December 2021.
\newblock URL \url{http://arxiv.org/abs/2111.06377}.
\newblock arXiv:2111.06377 [cs].

\bibitem[Mumford(1992)]{predictive_coding}
D~Mumford.
\newblock On the computational architecture of the neocortex.
\newblock \emph{Biol. Cybern.}, 66\penalty0 (3):\penalty0 241--251, January 1992.

\bibitem[Kogo and Trengove(2015)]{predictive_testable}
Naoki Kogo and Chris Trengove.
\newblock Is predictive coding theory articulated enough to be testable?
\newblock \emph{Front. Comput. Neurosci.}, 9:\penalty0 111, September 2015.

\bibitem[van~den Oord et~al.(2018)van~den Oord, Li, and Vinyals]{cont_pred}
Aaron van~den Oord, Yazhe Li, and Oriol Vinyals.
\newblock Representation learning with contrastive predictive coding.
\newblock July 2018.

\bibitem[Rao and Ballard(1999)]{predictive_cortex}
R~P Rao and D~H Ballard.
\newblock Predictive coding in the visual cortex: a functional interpretation of some extra-classical receptive-field effects.
\newblock \emph{Nat. Neurosci.}, 2\penalty0 (1):\penalty0 79--87, January 1999.

\bibitem[Palmer et~al.(2015)Palmer, Marre, Berry, and Bialek]{palmer_predictive_2015}
Stephanie~E. Palmer, Olivier Marre, Michael~J. Berry, and William Bialek.
\newblock Predictive information in a sensory population.
\newblock \emph{Proceedings of the National Academy of Sciences of the United States of America}, 112\penalty0 (22):\penalty0 6908--6913, June 2015.
\newblock ISSN 0027-8424.
\newblock \doi{10.1073/pnas.1506855112}.
\newblock URL \url{https://www.ncbi.nlm.nih.gov/pmc/articles/PMC4460449/}.

\bibitem[Friston(2005)]{predictive_friston}
Karl Friston.
\newblock A theory of cortical responses.
\newblock \emph{Philosophical Transactions of the Royal Society B: Biological Sciences}, 360\penalty0 (1456):\penalty0 815--836, April 2005.
\newblock \doi{10.1098/rstb.2005.1622}.
\newblock URL \url{https://royalsocietypublishing.org/doi/10.1098/rstb.2005.1622}.

\bibitem[Lotter et~al.(2017)Lotter, Kreiman, and Cox]{prednet}
William Lotter, Gabriel Kreiman, and David Cox.
\newblock Deep {Predictive} {Coding} {Networks} for {Video} {Prediction} and {Unsupervised} {Learning}, March 2017.
\newblock URL \url{http://arxiv.org/abs/1605.08104}.
\newblock arXiv:1605.08104 [cs].

\bibitem[Recanatesi et~al.(2020)Recanatesi, Farrell, Lajoie, Deneve, Rigotti, and Shea-Brown]{predictive_placecells}
Stefano Recanatesi, Matthew Farrell, Guillaume Lajoie, Sophie Deneve, Mattia Rigotti, and Eric Shea-Brown.
\newblock Predictive learning as a network mechanism for extracting low-dimensional latent space representations.
\newblock \emph{bioRxiv}, 2020.
\newblock \doi{10.1101/471987}.
\newblock URL \url{https://www.biorxiv.org/content/early/2020/09/17/471987}.

\bibitem[Catani and Thiebaut~de Schotten(2012)]{human_brain_atlas}
Marco Catani and Michel Thiebaut~de Schotten.
\newblock \emph{Atlas of Human Brain Connections}.
\newblock Oxford University Press, 03 2012.
\newblock ISBN 9780199541164.
\newblock \doi{10.1093/med/9780199541164.001.0001}.
\newblock URL \url{https://doi.org/10.1093/med/9780199541164.001.0001}.

\bibitem[Fan et~al.(2016)Fan, Li, Zhuo, Zhang, Wang, Chen, Yang, Chu, Xie, Laird, Fox, Eickhoff, Yu, and Jiang]{Fan2016_brainnetome}
Lingzhong Fan, Hai Li, Junjie Zhuo, Yu~Zhang, Jiaojian Wang, Liangfu Chen, Zhengyi Yang, Congying Chu, Sangma Xie, Angela~R. Laird, Peter~T. Fox, Simon~B. Eickhoff, Chunshui Yu, and Tianzi Jiang.
\newblock The human brainnetome atlas: A new brain atlas based on connectional architecture.
\newblock \emph{Cerebral Cortex}, 26\penalty0 (8):\penalty0 3508--3526, 07 2016.
\newblock ISSN 1047-3211.
\newblock \doi{10.1093/cercor/bhw157}.
\newblock URL \url{https://doi.org/10.1093/cercor/bhw157}.

\bibitem[Bullmore and Sporns(2009)]{Bullmore2009_graphtheory}
Edward Bullmore and Olaf Sporns.
\newblock Complex brain networks: Graph theoretical analysis of structural and functional systems.
\newblock \emph{Nature reviews. Neuroscience}, 10:\penalty0 186--98, 03 2009.
\newblock \doi{10.1038/nrn2575}.

\bibitem[Joshi et~al.(2022)Joshi, Choi, Liu, Chong, Sonkar, Gonzalez-Martinez, Nair, Wisnowski, Haldar, Shattuck, Damasio, and Leahy]{Joshi2022_fmriatlas}
Anand~A. Joshi, Soyoung Choi, Yijun Liu, Minqi Chong, Gaurav Sonkar, Jorge Gonzalez-Martinez, Dileep Nair, Jessica~L. Wisnowski, Justin~P. Haldar, David~W. Shattuck, Hanna Damasio, and Richard~M. Leahy.
\newblock A hybrid high-resolution anatomical mri atlas with sub-parcellation of cortical gyri using resting fmri.
\newblock \emph{Journal of Neuroscience Methods}, 374:\penalty0 109566, 2022.
\newblock ISSN 0165-0270.
\newblock \doi{https://doi.org/10.1016/j.jneumeth.2022.109566}.
\newblock URL \url{https://www.sciencedirect.com/science/article/pii/S0165027022000930}.

\bibitem[Boucsein et~al.(2011)Boucsein, Nawrot, Schnepel, and Aertsen]{boucsein_beyond_2011}
Clemens Boucsein, Martin Nawrot, Philipp Schnepel, and Ad~Aertsen.
\newblock Beyond the {Cortical} {Column}: {Abundance} and {Physiology} of {Horizontal} {Connections} {Imply} a {Strong} {Role} for {Inputs} from the {Surround}.
\newblock \emph{Frontiers in Neuroscience}, 5, April 2011.
\newblock ISSN 1662-453X.
\newblock \doi{10.3389/fnins.2011.00032}.
\newblock URL \url{https://www.frontiersin.org/journals/neuroscience/articles/10.3389/fnins.2011.00032/full}.
\newblock Publisher: Frontiers.

\bibitem[Dreyer et~al.(2025)Dreyer, Haluts, Korman, Gov, Fonio, and Feinerman]{ants_humans}
Tabea Dreyer, Amir Haluts, Amos Korman, Nir Gov, Ehud Fonio, and Ofer Feinerman.
\newblock Comparing cooperative geometric puzzle solving in ants versus humans.
\newblock \emph{Proceedings of the National Academy of Sciences}, 122\penalty0 (1):\penalty0 e2414274121, 2025.
\newblock \doi{10.1073/pnas.2414274121}.
\newblock URL \url{https://www.pnas.org/doi/abs/10.1073/pnas.2414274121}.

\bibitem[Feinerman et~al.(2018)Feinerman, Pinkoviezky, Gelblum, Fonio, and Gov]{ants_physics}
Ofer Feinerman, Itai Pinkoviezky, Aviram Gelblum, Ehud Fonio, and Nir~S. Gov.
\newblock The physics of cooperative transport in groups of ants.
\newblock \emph{Nature Physics}, 14\penalty0 (7):\penalty0 683--693, July 2018.
\newblock ISSN 1745-2481.
\newblock \doi{10.1038/s41567-018-0107-y}.
\newblock URL \url{https://www.nature.com/articles/s41567-018-0107-y}.

\bibitem[Katz et~al.(2011)Katz, Tunstrøm, Ioannou, Huepe, and Couzin]{schooling_fish}
Yael Katz, Kolbjørn Tunstrøm, Christos~C. Ioannou, Cristián Huepe, and Iain~D. Couzin.
\newblock Inferring the structure and dynamics of interactions in schooling fish.
\newblock \emph{Proceedings of the National Academy of Sciences}, 108\penalty0 (46):\penalty0 18720--18725, November 2011.
\newblock \doi{10.1073/pnas.1107583108}.
\newblock URL \url{https://www.pnas.org/doi/10.1073/pnas.1107583108}.
\newblock Publisher: Proceedings of the National Academy of Sciences.

\bibitem[Dkhili et~al.(2017)Dkhili, Berger, Idrissi~Hassani, Ghaout, Peters, and Piou]{locusts_dkhili_self-organized_2017}
Jamila Dkhili, Uta Berger, Lalla~Mina Idrissi~Hassani, Saïd Ghaout, Ronny Peters, and Cyril Piou.
\newblock Self-organized spatial structures of locust groups emerging from local interaction.
\newblock \emph{Ecological Modelling}, 361:\penalty0 26--40, October 2017.
\newblock ISSN 0304-3800.
\newblock \doi{10.1016/j.ecolmodel.2017.07.020}.
\newblock URL \url{https://www.sciencedirect.com/science/article/pii/S0304380017303423}.

\bibitem[Reynolds(1987)]{boids_reynolds1987flocks}
Craig~W Reynolds.
\newblock Flocks, herds and schools: A distributed behavioral model.
\newblock In \emph{Proceedings of the 14th annual conference on Computer graphics and interactive techniques}, pages 25--34, 1987.

\bibitem[Karpas et~al.(2017)Karpas, Shklarsh, and Schneidman]{socialtaxis}
Ehud~D Karpas, Adi Shklarsh, and Elad Schneidman.
\newblock Information socialtaxis and efficient collective behavior emerging in groups of information-seeking agents.
\newblock \emph{Proc. Natl. Acad. Sci. U. S. A.}, 114\penalty0 (22):\penalty0 5589--5594, May 2017.

\bibitem[Brazowski and Schneidman(2020)]{altruistic}
Benjamin Brazowski and Elad Schneidman.
\newblock Collective learning by ensembles of altruistic diversifying neural networks, June 2020.

\bibitem[Bell and Bala(2015)]{siamese}
Sean Bell and Kavita Bala.
\newblock Learning visual similarity for product design with convolutional neural networks.
\newblock \emph{ACM Transactions on Graphics (TOG)}, 34:\penalty0 1 -- 10, 2015.
\newblock URL \url{https://api.semanticscholar.org/CorpusID:5518345}.

\bibitem[Chopra et~al.(2005{\natexlab{a}})Chopra, Hadsell, and LeCun]{siamese_lecun}
S.~Chopra, R.~Hadsell, and Y.~LeCun.
\newblock Learning a similarity metric discriminatively, with application to face verification.
\newblock In \emph{2005 IEEE Computer Society Conference on Computer Vision and Pattern Recognition (CVPR'05)}, volume~1, pages 539--546 vol. 1, 2005{\natexlab{a}}.
\newblock \doi{10.1109/CVPR.2005.202}.

\bibitem[Bromley et~al.(1993)Bromley, Guyon, LeCun, S\"{a}ckinger, and Shah]{siamese_sign}
Jane Bromley, Isabelle Guyon, Yann LeCun, Eduard S\"{a}ckinger, and Roopak Shah.
\newblock Signature verification using a "siamese" time delay neural network.
\newblock In J.~Cowan, G.~Tesauro, and J.~Alspector, editors, \emph{Advances in Neural Information Processing Systems}, volume~6. Morgan-Kaufmann, 1993.
\newblock URL \url{https://proceedings.neurips.cc/paper_files/paper/1993/file/288cc0ff022877bd3df94bc9360b9c5d-Paper.pdf}.

\bibitem[Schroff et~al.(2015)Schroff, Kalenichenko, and Philbin]{triplet}
Florian Schroff, Dmitry Kalenichenko, and James Philbin.
\newblock Facenet: A unified embedding for face recognition and clustering.
\newblock In \emph{2015 IEEE Conference on Computer Vision and Pattern Recognition (CVPR)}, page 815–823. IEEE, June 2015.
\newblock \doi{10.1109/cvpr.2015.7298682}.
\newblock URL \url{http://dx.doi.org/10.1109/CVPR.2015.7298682}.

\bibitem[Chen et~al.(2020)Chen, Kornblith, Norouzi, and Hinton]{simclr}
Ting Chen, Simon Kornblith, Mohammad Norouzi, and Geoffrey Hinton.
\newblock A simple framework for contrastive learning of visual representations, February 2020.

\bibitem[Arora et~al.(2019)Arora, Khandeparkar, Khodak, Plevrakis, and Saunshi]{contrastive}
Sanjeev Arora, Hrishikesh Khandeparkar, Mikhail Khodak, Orestis Plevrakis, and Nikunj Saunshi.
\newblock A theoretical analysis of contrastive unsupervised representation learning.
\newblock 2019.

\bibitem[Yerxa et~al.(2023)Yerxa, Kuang, Simoncelli, and Chung]{Yerxa2023_maxmanifoldcapacity}
Thomas Yerxa, Yilun Kuang, Eero Simoncelli, and SueYeon Chung.
\newblock Learning efficient coding of natural images with maximum manifold capacity representations.
\newblock In A.~Oh, T.~Naumann, A.~Globerson, K.~Saenko, M.~Hardt, and S.~Levine, editors, \emph{Advances in Neural Information Processing Systems}, volume~36, pages 24103--24128. Curran Associates, Inc., 2023.
\newblock URL \url{https://proceedings.neurips.cc/paper_files/paper/2023/file/4bc6e94f2308c888fb69626138a2633e-Paper-Conference.pdf}.

\bibitem[Bardes et~al.(2022)Bardes, Ponce, and LeCun]{vicreg}
Adrien Bardes, Jean Ponce, and Yann LeCun.
\newblock Vicreg: Variance-invariance-covariance regularization for self-supervised learning, 2022.
\newblock URL \url{https://arxiv.org/abs/2105.04906}.

\bibitem[Grill et~al.(2020)Grill, Strub, Altch{\'e}, Tallec, Richemond, Buchatskaya, Doersch, Pires, Guo, Azar, Piot, Kavukcuoglu, Munos, and Valko]{byol}
Jean-Bastien Grill, Florian Strub, Florent Altch{\'e}, Corentin Tallec, Pierre~H Richemond, Elena Buchatskaya, Carl Doersch, Bernardo~Avila Pires, Zhaohan~Daniel Guo, Mohammad~Gheshlaghi Azar, Bilal Piot, Koray Kavukcuoglu, R{\'e}mi Munos, and Michal Valko.
\newblock Bootstrap your own latent: A new approach to self-supervised learning.
\newblock 2020.

\bibitem[Zhuo et~al.(2023)Zhuo, Wang, Ma, and Wang]{noncontrastive_theo}
Zhijian Zhuo, Yifei Wang, Jinwen Ma, and Yisen Wang.
\newblock Towards a unified theoretical understanding of non-contrastive learning via rank differential mechanism, 2023.
\newblock URL \url{https://arxiv.org/abs/2303.02387}.

\bibitem[Caron et~al.(2021)Caron, Touvron, Misra, J{\'e}gou, Mairal, Bojanowski, and Joulin]{dino}
Mathilde Caron, Hugo Touvron, Ishan Misra, Herv{\'e} J{\'e}gou, Julien Mairal, Piotr Bojanowski, and Armand Joulin.
\newblock Emerging properties in self-supervised vision transformers.
\newblock April 2021.

\bibitem[Zhou et~al.(2021)Zhou, Wei, Wang, Shen, Xie, Yuille, and Kong]{ibot}
Jinghao Zhou, Chen Wei, Huiyu Wang, Wei Shen, Cihang Xie, Alan Yuille, and Tao Kong.
\newblock {iBOT}: Image {BERT} {Pre-Training} with online tokenizer.
\newblock November 2021.

\bibitem[Oquab et~al.(2023)Oquab, Darcet, Moutakanni, Vo, Szafraniec, Khalidov, Fernandez, Haziza, Massa, El-Nouby, Assran, Ballas, Galuba, Howes, Huang, Li, Misra, Rabbat, Sharma, Synnaeve, Xu, Jegou, Mairal, Labatut, Joulin, and Bojanowski]{dinov2}
Maxime Oquab, Timothée Darcet, Théo Moutakanni, Huy Vo, Marc Szafraniec, Vasil Khalidov, Pierre Fernandez, Daniel Haziza, Francisco Massa, Alaaeldin El-Nouby, Mahmoud Assran, Nicolas Ballas, Wojciech Galuba, Russell Howes, Po-Yao Huang, Shang-Wen Li, Ishan Misra, Michael Rabbat, Vasu Sharma, Gabriel Synnaeve, Hu~Xu, Hervé Jegou, Julien Mairal, Patrick Labatut, Armand Joulin, and Piotr Bojanowski.
\newblock {DINOv2}: {Learning} {Robust} {Visual} {Features} without {Supervision}, April 2023.
\newblock URL \url{http://arxiv.org/abs/2304.07193}.
\newblock arXiv:2304.07193 [cs].

\bibitem[Assran et~al.(2023)Assran, Duval, Misra, Bojanowski, Vincent, Rabbat, LeCun, and Ballas]{ijepa}
Mahmoud Assran, Quentin Duval, Ishan Misra, Piotr Bojanowski, Pascal Vincent, Michael Rabbat, Yann LeCun, and Nicolas Ballas.
\newblock Self-supervised learning from images with a joint-embedding predictive architecture, 2023.
\newblock URL \url{https://arxiv.org/abs/2301.08243}.

\bibitem[Siméoni et~al.(2025)Siméoni, Vo, Seitzer, Baldassarre, Oquab, Jose, Khalidov, Szafraniec, Yi, Ramamonjisoa, Massa, Haziza, Wehrstedt, Wang, Darcet, Moutakanni, Sentana, Roberts, Vedaldi, Tolan, Brandt, Couprie, Mairal, Jégou, Labatut, and Bojanowski]{dinov3}
Oriane Siméoni, Huy~V. Vo, Maximilian Seitzer, Federico Baldassarre, Maxime Oquab, Cijo Jose, Vasil Khalidov, Marc Szafraniec, Seungeun Yi, Michaël Ramamonjisoa, Francisco Massa, Daniel Haziza, Luca Wehrstedt, Jianyuan Wang, Timothée Darcet, Théo Moutakanni, Leonel Sentana, Claire Roberts, Andrea Vedaldi, Jamie Tolan, John Brandt, Camille Couprie, Julien Mairal, Hervé Jégou, Patrick Labatut, and Piotr Bojanowski.
\newblock Dinov3, 2025.
\newblock URL \url{https://arxiv.org/abs/2508.10104}.

\bibitem[Van~den Bergh et~al.(2010)Van~den Bergh, Zhang, Arckens, and Chino]{receptive_field}
Gert Van~den Bergh, Bin Zhang, Lutgarde Arckens, and Yuzo~M. Chino.
\newblock Receptive-field properties of {V1} and {V2} neurons in mice and macaque monkeys.
\newblock \emph{The Journal of Comparative Neurology}, 518\penalty0 (11):\penalty0 2051--2070, June 2010.
\newblock ISSN 1096-9861.
\newblock \doi{10.1002/cne.22321}.

\bibitem[Chopra et~al.(2005{\natexlab{b}})Chopra, Hadsell, and Lecun]{siamese_face}
Sumit Chopra, Raia Hadsell, and Yann Lecun.
\newblock Learning a similarity metric discriminatively, with application to face verification.
\newblock volume~1, pages 539-- 546 vol. 1, 07 2005{\natexlab{b}}.
\newblock ISBN 0-7695-2372-2.
\newblock \doi{10.1109/CVPR.2005.202}.

\bibitem[Krizhevsky(2012)]{cifar10}
Alex Krizhevsky.
\newblock Learning {Multiple} {Layers} of {Features} from {Tiny} {Images}.
\newblock \emph{University of Toronto}, May 2012.

\bibitem[Krizhevsky et~al.(2012)Krizhevsky, Sutskever, and Hinton]{alexnet}
Alex Krizhevsky, Ilya Sutskever, and Geoffrey~E Hinton.
\newblock Imagenet classification with deep convolutional neural networks.
\newblock In F.~Pereira, C.J. Burges, L.~Bottou, and K.Q. Weinberger, editors, \emph{Advances in Neural Information Processing Systems}, volume~25. Curran Associates, Inc., 2012.
\newblock URL \url{https://proceedings.neurips.cc/paper_files/paper/2012/file/c399862d3b9d6b76c8436e924a68c45b-Paper.pdf}.

\bibitem[Rosenblatt(1958)]{rosenblatt_mlp}
Frank Rosenblatt.
\newblock The perceptron: a probabilistic model for information storage and organization in the brain.
\newblock \emph{Psychological review}, 65 6:\penalty0 386--408, 1958.
\newblock URL \url{https://api.semanticscholar.org/CorpusID:12781225}.

\bibitem[Rumelhart et~al.(1986)Rumelhart, Hinton, and Williams]{backprop_mlp}
David~E. Rumelhart, Geoffrey~E. Hinton, and Ronald~J. Williams.
\newblock Learning representations by back-propagating errors.
\newblock \emph{Nature}, 323:\penalty0 533--536, 1986.
\newblock URL \url{https://api.semanticscholar.org/CorpusID:205001834}.

\bibitem[Hendrycks and Gimpel(2023)]{gelu}
Dan Hendrycks and Kevin Gimpel.
\newblock Gaussian {Error} {Linear} {Units} ({GELUs}), June 2023.
\newblock URL \url{http://arxiv.org/abs/1606.08415}.
\newblock arXiv:1606.08415 [cs].

\bibitem[Dozat(2016)]{nadam}
Timothy Dozat.
\newblock Incorporating {Nesterov} {Momentum} into {Adam}.
\newblock February 2016.
\newblock URL \url{https://openreview.net/forum?id=OM0jvwB8jIp57ZJjtNEZ}.

\bibitem[Maoz et~al.(2020)Maoz, Tkačik, Esteki, Kiani, and Schneidman]{Maoz2020_RP}
Ori Maoz, Gašper Tkačik, Mohamad~Saleh Esteki, Roozbeh Kiani, and Elad Schneidman.
\newblock Learning probabilistic neural representations with randomly connected circuits.
\newblock \emph{Proceedings of the National Academy of Sciences}, 117\penalty0 (40):\penalty0 25066--25073, 2020.
\newblock \doi{10.1073/pnas.1912804117}.
\newblock URL \url{https://www.pnas.org/doi/abs/10.1073/pnas.1912804117}.

\bibitem[Jones and Palmer(1987)]{receptive_field_multiple}
J.~P. Jones and L.~A. Palmer.
\newblock The two-dimensional spatial structure of simple receptive fields in cat striate cortex.
\newblock \emph{Journal of Neurophysiology}, 58\penalty0 (6):\penalty0 1187--1211, 1987.
\newblock \doi{10.1152/jn.1987.58.6.1187}.
\newblock URL \url{https://doi.org/10.1152/jn.1987.58.6.1187}.
\newblock PMID: 3437330.

\bibitem[Siegle et~al.(2021)Siegle, Jia, Durand, Gale, Bennett, Graddis, Heller, Ramirez, Choi, Luviano, Groblewski, Ahmed, Arkhipov, Bernard, Billeh, Brown, Buice, Cain, Caldejon, Casal, Cho, Chvilicek, Cox, Dai, Denman, De~Vries, Dietzman, Esposito, Farrell, Feng, Galbraith, Garrett, Gelfand, Hancock, Harris, Howard, Hu, Hytnen, Iyer, Jessett, Johnson, Kato, Kiggins, Lambert, Lecoq, Ledochowitsch, Lee, Leon, Li, Liang, Long, Mace, Melchior, Millman, Mollenkopf, Nayan, Ng, Ngo, Nguyen, Nicovich, North, Ocker, Ollerenshaw, Oliver, Pachitariu, Perkins, Reding, Reid, Robertson, Ronellenfitch, Seid, Slaughterbeck, Stoecklin, Sullivan, Sutton, Swapp, Thompson, Turner, Wakeman, Whitesell, Williams, Williford, Young, Zeng, Naylor, Phillips, Reid, Mihalas, Olsen, and Koch]{allen_data}
Joshua~H. Siegle, Xiaoxuan Jia, Séverine Durand, Sam Gale, Corbett Bennett, Nile Graddis, Greggory Heller, Tamina~K. Ramirez, Hannah Choi, Jennifer~A. Luviano, Peter~A. Groblewski, Ruweida Ahmed, Anton Arkhipov, Amy Bernard, Yazan~N. Billeh, Dillan Brown, Michael~A. Buice, Nicolas Cain, Shiella Caldejon, Linzy Casal, Andrew Cho, Maggie Chvilicek, Timothy~C. Cox, Kael Dai, Daniel~J. Denman, Saskia E.~J. De~Vries, Roald Dietzman, Luke Esposito, Colin Farrell, David Feng, John Galbraith, Marina Garrett, Emily~C. Gelfand, Nicole Hancock, Julie~A. Harris, Robert Howard, Brian Hu, Ross Hytnen, Ramakrishnan Iyer, Erika Jessett, Katelyn Johnson, India Kato, Justin Kiggins, Sophie Lambert, Jerome Lecoq, Peter Ledochowitsch, Jung~Hoon Lee, Arielle Leon, Yang Li, Elizabeth Liang, Fuhui Long, Kyla Mace, Jose Melchior, Daniel Millman, Tyler Mollenkopf, Chelsea Nayan, Lydia Ng, Kiet Ngo, Thuyahn Nguyen, Philip~R. Nicovich, Kat North, Gabriel~Koch Ocker, Doug Ollerenshaw, Michael Oliver, Marius Pachitariu, Jed Perkins,
  Melissa Reding, David Reid, Miranda Robertson, Kara Ronellenfitch, Sam Seid, Cliff Slaughterbeck, Michelle Stoecklin, David Sullivan, Ben Sutton, Jackie Swapp, Carol Thompson, Kristen Turner, Wayne Wakeman, Jennifer~D. Whitesell, Derric Williams, Ali Williford, Rob Young, Hongkui Zeng, Sarah Naylor, John~W. Phillips, R.~Clay Reid, Stefan Mihalas, Shawn~R. Olsen, and Christof Koch.
\newblock Survey of spiking in the mouse visual system reveals functional hierarchy.
\newblock \emph{Nature}, 592\penalty0 (7852):\penalty0 86--92, April 2021.
\newblock ISSN 0028-0836, 1476-4687.
\newblock \doi{10.1038/s41586-020-03171-x}.
\newblock URL \url{https://www.nature.com/articles/s41586-020-03171-x}.

\bibitem[Richert et~al.(2013)Richert, Albright, and Krekelberg]{receptive_field_mt}
Micah Richert, Thomas~D. Albright, and Bart Krekelberg.
\newblock The complex structure of receptive fields in the middle temporal area.
\newblock \emph{Frontiers in Systems Neuroscience}, Volume 7 - 2013, 2013.
\newblock ISSN 1662-5137.
\newblock \doi{10.3389/fnsys.2013.00002}.
\newblock URL \url{https://www.frontiersin.org/journals/systems-neuroscience/articles/10.3389/fnsys.2013.00002}.

\bibitem[Dosovitskiy et~al.(2021)Dosovitskiy, Beyer, Kolesnikov, Weissenborn, Zhai, Unterthiner, Dehghani, Minderer, Heigold, Gelly, Uszkoreit, and Houlsby]{vit}
Alexey Dosovitskiy, Lucas Beyer, Alexander Kolesnikov, Dirk Weissenborn, Xiaohua Zhai, Thomas Unterthiner, Mostafa Dehghani, Matthias Minderer, Georg Heigold, Sylvain Gelly, Jakob Uszkoreit, and Neil Houlsby.
\newblock An {Image} is {Worth} 16x16 {Words}: {Transformers} for {Image} {Recognition} at {Scale}, June 2021.
\newblock URL \url{http://arxiv.org/abs/2010.11929}.
\newblock arXiv:2010.11929 [cs].

\bibitem[Vaswani et~al.(2023)Vaswani, Shazeer, Parmar, Uszkoreit, Jones, Gomez, Kaiser, and Polosukhin]{attn}
Ashish Vaswani, Noam Shazeer, Niki Parmar, Jakob Uszkoreit, Llion Jones, Aidan~N. Gomez, Lukasz Kaiser, and Illia Polosukhin.
\newblock Attention {Is} {All} {You} {Need}, August 2023.
\newblock URL \url{http://arxiv.org/abs/1706.03762}.
\newblock arXiv:1706.03762 [cs].

\bibitem[Hassani et~al.(2022)Hassani, Walton, Shah, Abuduweili, Li, and Shi]{vitlite}
Ali Hassani, Steven Walton, Nikhil Shah, Abulikemu Abuduweili, Jiachen Li, and Humphrey Shi.
\newblock Escaping the {Big} {Data} {Paradigm} with {Compact} {Transformers}, June 2022.
\newblock URL \url{http://arxiv.org/abs/2104.05704}.
\newblock arXiv:2104.05704 [cs].

\bibitem[Ellwood(2024)]{hebattn}
Ian~T. Ellwood.
\newblock Short-term {Hebbian} learning can implement transformer-like attention.
\newblock \emph{PLoS computational biology}, 20\penalty0 (1):\penalty0 e1011843, January 2024.
\newblock ISSN 1553-7358.
\newblock \doi{10.1371/journal.pcbi.1011843}.

\bibitem[Kozachkov et~al.(2023)Kozachkov, Kastanenka, and Krotov]{astroattn}
Leo Kozachkov, Ksenia~V. Kastanenka, and Dmitry Krotov.
\newblock Building transformers from neurons and astrocytes.
\newblock \emph{Proceedings of the National Academy of Sciences}, 120\penalty0 (34):\penalty0 e2219150120, August 2023.
\newblock \doi{10.1073/pnas.2219150120}.
\newblock URL \url{https://www.pnas.org/doi/10.1073/pnas.2219150120}.
\newblock Publisher: Proceedings of the National Academy of Sciences.

\bibitem[Wiskott and Sejnowski(2002)]{sfa}
Laurenz Wiskott and Terrence~J. Sejnowski.
\newblock Slow {Feature} {Analysis}: {Unsupervised} {Learning} of {Invariances}.
\newblock \emph{Neural Computation}, 14\penalty0 (4):\penalty0 715--770, April 2002.
\newblock ISSN 0899-7667, 1530-888X.
\newblock \doi{10.1162/089976602317318938}.
\newblock URL \url{https://direct.mit.edu/neco/article/14/4/715-770/6583}.

\bibitem[Halvagal and Zenke(2023)]{lpl}
Manu~Srinath Halvagal and Friedemann Zenke.
\newblock The combination of {Hebbian} and predictive plasticity learns invariant object representations in deep sensory networks.
\newblock \emph{Nature Neuroscience}, 26\penalty0 (11):\penalty0 1906--1915, November 2023.
\newblock ISSN 1546-1726.
\newblock \doi{10.1038/s41593-023-01460-y}.
\newblock URL \url{https://www.nature.com/articles/s41593-023-01460-y}.
\newblock Publisher: Nature Publishing Group.

\bibitem[Bienenstock et~al.(1982)Bienenstock, Cooper, and Munro]{bcm}
Elie~L Bienenstock, Leon~N Cooper, and Paul~W Munro.
\newblock Theory for the development of neuron selectivity: orientation specificity and binocular interaction in visual cortex.
\newblock \emph{Journal of Neuroscience}, 2\penalty0 (1):\penalty0 32--48, 1982.

\bibitem[Schrimpf et~al.(2020)Schrimpf, Kubilius, Hong, Majaj, Rajalingham, Issa, Kar, Bashivan, Prescott-Roy, Geiger, Schmidt, Yamins, and DiCarlo]{brainscore}
Martin Schrimpf, Jonas Kubilius, Ha~Hong, Najib~J. Majaj, Rishi Rajalingham, Elias~B. Issa, Kohitij Kar, Pouya Bashivan, Jonathan Prescott-Roy, Franziska Geiger, Kailyn Schmidt, Daniel L.~K. Yamins, and James~J. DiCarlo.
\newblock Brain-score: Which artificial neural network for object recognition is most brain-like?
\newblock \emph{bioRxiv}, 2020.
\newblock \doi{10.1101/407007}.
\newblock URL \url{https://www.biorxiv.org/content/early/2020/01/02/407007}.

\bibitem[Zhuang et~al.(2021)Zhuang, Yan, Nayebi, Schrimpf, Frank, DiCarlo, and Yamins]{brainscore_unsupervised}
Chengxu Zhuang, Siming Yan, Aran Nayebi, Martin Schrimpf, Michael~C. Frank, James~J. DiCarlo, and Daniel L.~K. Yamins.
\newblock Unsupervised neural network models of the ventral visual stream.
\newblock \emph{Proceedings of the National Academy of Sciences}, 118\penalty0 (3):\penalty0 e2014196118, 2021.
\newblock \doi{10.1073/pnas.2014196118}.
\newblock URL \url{https://www.pnas.org/doi/abs/10.1073/pnas.2014196118}.

\bibitem[Margalit et~al.(2024)Margalit, Lee, Finzi, DiCarlo, Grill-Spector, and Yamins]{tdann}
Eshed Margalit, Hyodong Lee, Dawn Finzi, James~J. DiCarlo, Kalanit Grill-Spector, and Daniel~L.K. Yamins.
\newblock A unifying framework for functional organization in early and higher ventral visual cortex.
\newblock \emph{Neuron}, 112\penalty0 (14):\penalty0 2435--2451.e7, 2024.
\newblock ISSN 0896-6273.
\newblock \doi{https://doi.org/10.1016/j.neuron.2024.04.018}.
\newblock URL \url{https://www.sciencedirect.com/science/article/pii/S0896627324002794}.

\bibitem[Schneider et~al.(2023)Schneider, Lee, and Mathis]{cebra}
Steffen Schneider, Jin~Hwa Lee, and Mackenzie~Weygandt Mathis.
\newblock Learnable latent embeddings for joint behavioural and neural analysis.
\newblock \emph{Nature}, 617\penalty0 (7960):\penalty0 360--368, May 2023.
\newblock ISSN 1476-4687.
\newblock \doi{10.1038/s41586-023-06031-6}.
\newblock URL \url{https://www.nature.com/articles/s41586-023-06031-6}.
\newblock Number: 7960 Publisher: Nature Publishing Group.

\bibitem[Brodmann(1909)]{brodmann}
K.~Brodmann.
\newblock \emph{Vergleichende Lokalisationslehre der Grosshirnrinde in ihren Prinzipien dargestellt auf Grund des Zellenbaues von Dr. K. Brodmann ...}
\newblock J.A. Barth, 1909.
\newblock URL \url{https://books.google.co.il/books?id=Qw5KQwAACAAJ}.

\bibitem[Haber et~al.(2023)Haber, Wanner, Friedrich, and Schneidman]{adam_structure}
Adam Haber, Adrian~A. Wanner, Rainer~W. Friedrich, and Elad Schneidman.
\newblock The structure and function of neural connectomes are shaped by a small number of design principles.
\newblock \emph{bioRxiv}, 2023.
\newblock \doi{10.1101/2023.03.15.532611}.
\newblock URL \url{https://www.biorxiv.org/content/early/2023/03/15/2023.03.15.532611}.

\bibitem[Allman and Kaas(1971)]{retinotopy}
John~M. Allman and Jon~H. Kaas.
\newblock Representation of the visual field in striate and adjoining cortex of the owl monkey(aotus trivirgatus).
\newblock \emph{Brain Research}, 35\penalty0 (1):\penalty0 89--106, 1971.
\newblock ISSN 0006-8993.
\newblock \doi{https://doi.org/10.1016/0006-8993(71)90596-8}.
\newblock URL \url{https://www.sciencedirect.com/science/article/pii/0006899371905968}.

\bibitem[Shapson-Coe et~al.(2024)Shapson-Coe, Januszewski, Berger, Pope, Wu, Blakely, Schalek, Li, Wang, Maitin-Shepard, Karlupia, Dorkenwald, Sjostedt, Leavitt, Lee, Troidl, Collman, Bailey, Fitzmaurice, Kar, Field, Wu, Wagner-Carena, Aley, Lau, Lin, Wei, Pfister, Peleg, Jain, and Lichtman]{human_connectome}
Alexander Shapson-Coe, Michał Januszewski, Daniel~R. Berger, Art Pope, Yuelong Wu, Tim Blakely, Richard~L. Schalek, Peter~H. Li, Shuohong Wang, Jeremy Maitin-Shepard, Neha Karlupia, Sven Dorkenwald, Evelina Sjostedt, Laramie Leavitt, Dongil Lee, Jakob Troidl, Forrest Collman, Luke Bailey, Angerica Fitzmaurice, Rohin Kar, Benjamin Field, Hank Wu, Julian Wagner-Carena, David Aley, Joanna Lau, Zudi Lin, Donglai Wei, Hanspeter Pfister, Adi Peleg, Viren Jain, and Jeff~W. Lichtman.
\newblock A petavoxel fragment of human cerebral cortex reconstructed at nanoscale resolution.
\newblock \emph{Science}, 384\penalty0 (6696):\penalty0 eadk4858, 2024.
\newblock \doi{10.1126/science.adk4858}.
\newblock URL \url{https://www.science.org/doi/abs/10.1126/science.adk4858}.

\bibitem[Bosking et~al.(1997)Bosking, Zhang, Schofield, and Fitzpatrick]{orientation_maps}
William~H. Bosking, Ying Zhang, Brett~R. Schofield, and David Fitzpatrick.
\newblock Orientation selectivity and the arrangement of horizontal connections in tree shrew striate cortex.
\newblock \emph{The Journal of Neuroscience}, 17:\penalty0 2112 -- 2127, 1997.
\newblock URL \url{https://api.semanticscholar.org/CorpusID:15271847}.

\bibitem[Deb et~al.(2025)Deb, Deb, and Murty]{toponet}
Mayukh Deb, Mainak Deb, and N.~Apurva~Ratan Murty.
\newblock Toponets: High performing vision and language models with brain-like topography, 2025.
\newblock URL \url{https://arxiv.org/abs/2501.16396}.

\bibitem[Kingma and Ba(2017)]{adam}
Diederik~P. Kingma and Jimmy Ba.
\newblock Adam: {A} {Method} for {Stochastic} {Optimization}, January 2017.
\newblock URL \url{http://arxiv.org/abs/1412.6980}.
\newblock arXiv:1412.6980 [cs].

\bibitem[Nesterov(1983)]{nesterov}
Y.~Nesterov.
\newblock A method for solving the convex programming problem with convergence rate \(o(1/k^2)\), 1983.
\newblock URL \url{https://cir.nii.ac.jp/crid/1370862715914709505}.

\end{thebibliography}



\clearpage
\newpage
\section*{Supplementary information}



\setcounter{figure}{0}
\setcounter{table}{0}
\setcounter{section}{0}
\renewcommand{\thefigure}{S\arabic{figure}}
\renewcommand{\thetable}{S\arabic{table}}
\renewcommand{\thesection}{S\arabic{section}}

\subsection*{Decomposition of \(L_{CLoSeR}\) and its probabilistic interpretation}\label{Sup: Further intuition for the loss}

To understand how minimizing \(L_{CLoSeR}\) (equation \ref{eq:L_CLoSeR}) affects the encoders, we further develop its formula to an equivalent loss, by substituting \(p(\mu_j(x_b)|\mu_i(x_b))\) according to its definition in equation \ref{eq:likelihood}:

\begin{equation}\label{eq:L_CLoSeR_develop}
\begin{split}
    L_{CLoSeR} & = \langle-\log(\frac{\psi_{ij}(x_b,x_k)}{\sum_{m=1}^B \psi_{ij}(x_b,x_m)})\rangle_{i\neq j, b} \\
                & = \langle-\log(\psi_{ij}(x_b,x_k))+\log({\sum_{m=1}^B \psi_{ij}(x_b,x_m)})\rangle_{i\neq j, b}
\end{split}
\end{equation}

Next, we substitute \(\psi_{ij}(x_b,x_b)\) according to its definition in equation \ref{eq:similarity}:

\begin{equation}
\begin{split}
                L_{CLoSeR}& = \langle-\log(\exp(-\frac{d^2_{ij}(x_b,x_b)}{\tau}))+\log({\sum_{m=1}^B \psi_{ij}(x_b,x_m)})\rangle_{i\neq j, b}
                \\
                & = \langle\frac{d^2_{ij}(x_b,x_b)}{\tau} + \log(\sum_{m=1}^B \psi_{ij}(x_b,x_m))\rangle_{i,j\neq i, b}
\end{split}
\end{equation}

Thus, we minimize \(d_{ij}(x_b,x_b)\), thereby ``pulling''  the encoders towards each other on the same stimuli, and minimize the overall similarity with the representations of other samples \(\psi_{ij}(x_b,x_m)\). This creates a ``contrast'' in the embedding space between samples that came from the same images and embeddings of different ones. 

Another interpretation of the loss function we use (inspired by \nociteclickt{cont_pred}) is that we maximize our model's ability to predict the masked input of one encoder from the masked input of the other encoders: denoting the masking of example \(b\) according to the \(i\)'th encoder's mask by \(m_i(x_b)\), we wish to estimate \(\mathbb{P}(m_j(x_b)|m_i(x_b))\). We define \(\psi_{ij}(x_b,x_k)\propto\mathbb{P}(m_j(x_b)|m_i(x_b))\) and use the negative examples to calculate \(p(\mu_j(x_k)|\mu_i(x_b))\), which approximates \(p(\mu_j(x_b)|\mu_i(x_k))\approx \mathbb{P}(m_j(x_k)|m_i(x_b))\). Finally, we maximize \(p(\mu_j(x_b)|\mu_i(x_b))\) to train our encoders.

\subsection*{A tradeoff between receptive field size and spreadingness}\label{Sup: A tradeoff between receptive field size and spreadingness}
\begin{figure*}
     \centering
     \includegraphics[width=\textwidth]{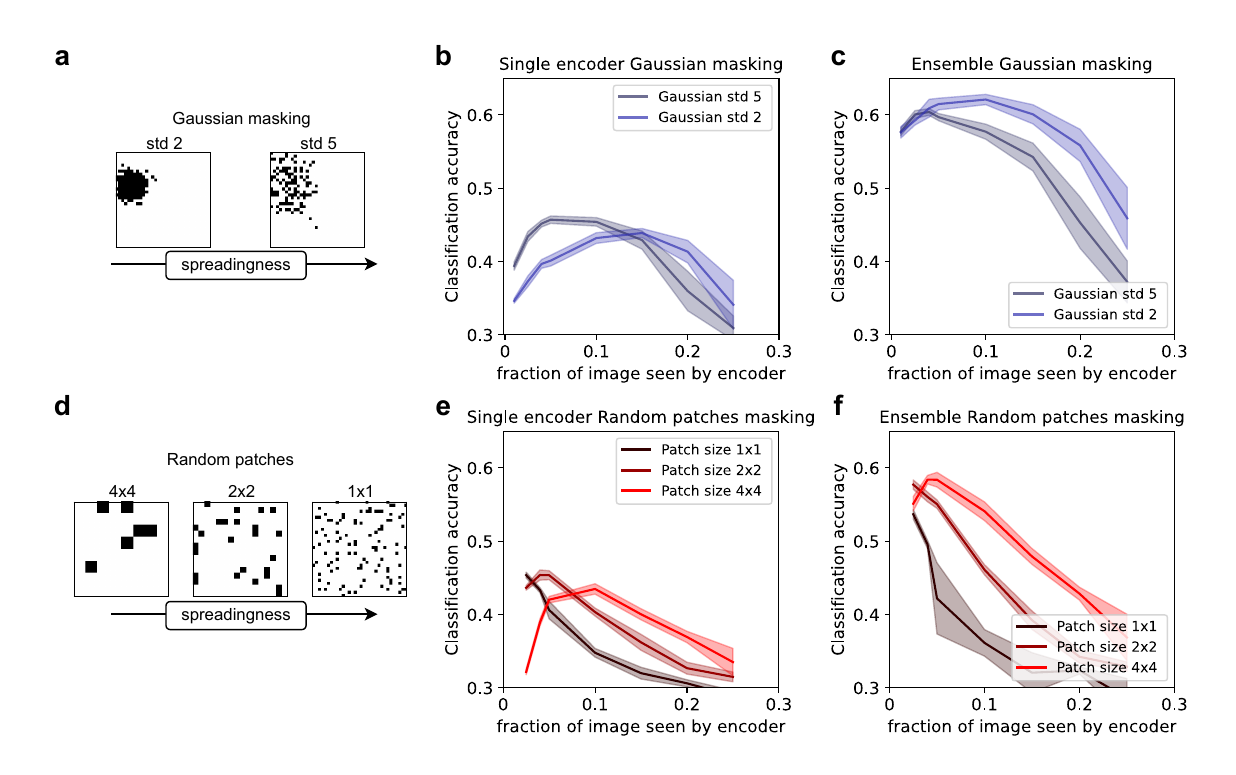}
     \caption{\textbf{A tradeoff between receptive field size and spreadingness:}
     \textbf{(a)} Example of the Gaussian masking method and how it changes as its spreadingness changes, parameterized by the standard deviation of the Gaussian. Both examples receive 10\% of the image pixels. 
     \textbf{(b)} The average accuracy of a supervised linear classifier decoding the 10 MLP encoders with Gaussian masking trained with CLoSeR as a function of the fraction of the image each encoder receives as input. Each line shows results for a different standard deviation when sampling the mask (negatively-correlated with locality), and the shading shows the 95\% confidence interval over different masking seeds (\(n=10\)). 
     \textbf{(c)} Same as \textbf{b} for ensemble accuracy.
     \textbf{(d)} Example of the random patches masking method and how it changes with the patch size. All examples receive 10\% of the image pixels.
     \textbf{(e-f)} Same as \textbf{b} and \textbf{c} for random patches masking and color corresponds to different patch-sizes (correlated with spreadingngess).
     }
    \label{fig_sup_spreadingness}
\end{figure*}

We have found in section \ref{res_vis_q} that when the encoders see ``too much'' of the external input, the goal of agreement between them becomes too easy and their performance goes down, as they agree on features of the stimulus that are not semantically interesting. We explored this notion further by controlling the spreadingness of the receptive field (see fig. \ref{fig_sup_spreadingness}a and d). For the Gaussian masking method, we increased the spreadingness by increasing the Gaussian's standard deviation (see Methods \nameref{Methods: Mask sampling}). For the random patches, we increased the spreadingness by making the patches smaller (thereby allowing more spreading for the same number of pixels). 
For both the Gaussian and the random patches masking, we see a tradeoff: as spreadingness grows, the optimal image fraction goes down (see fig. \ref{fig_sup_spreadingness}b and e). Also, the single encoder's accuracy at the optimal image fraction went up with the spreadingness. For the ensemble we see similar and opposite trends (see fig. \ref{fig_sup_spreadingness}c and f). In general, the ensemble required a smaller image fraction for optimal performance. Also, similarly to the single encoder, as spreadingness increased, the optimal image fraction went down, keeping this tradeoff. In contrast to the case of a single encoder, as spreadingness went up, the ensemble accuracy went down. In summary, this ``tradeoff'' between image fraction and spreadingness is evident in both single encoder and ensemble, and these parameters also control a different tradeoff between single encoder and ensemble accuracies. From a biological perspective, this tradeoff between image fraction and spreadingness is analogous to the energetic costs of having more synapses vs. longer axons. The second tradeoff between single encoder and ensemble is more functional in nature.


\begin{figure*}
     \centering
     \includegraphics[width=\textwidth]{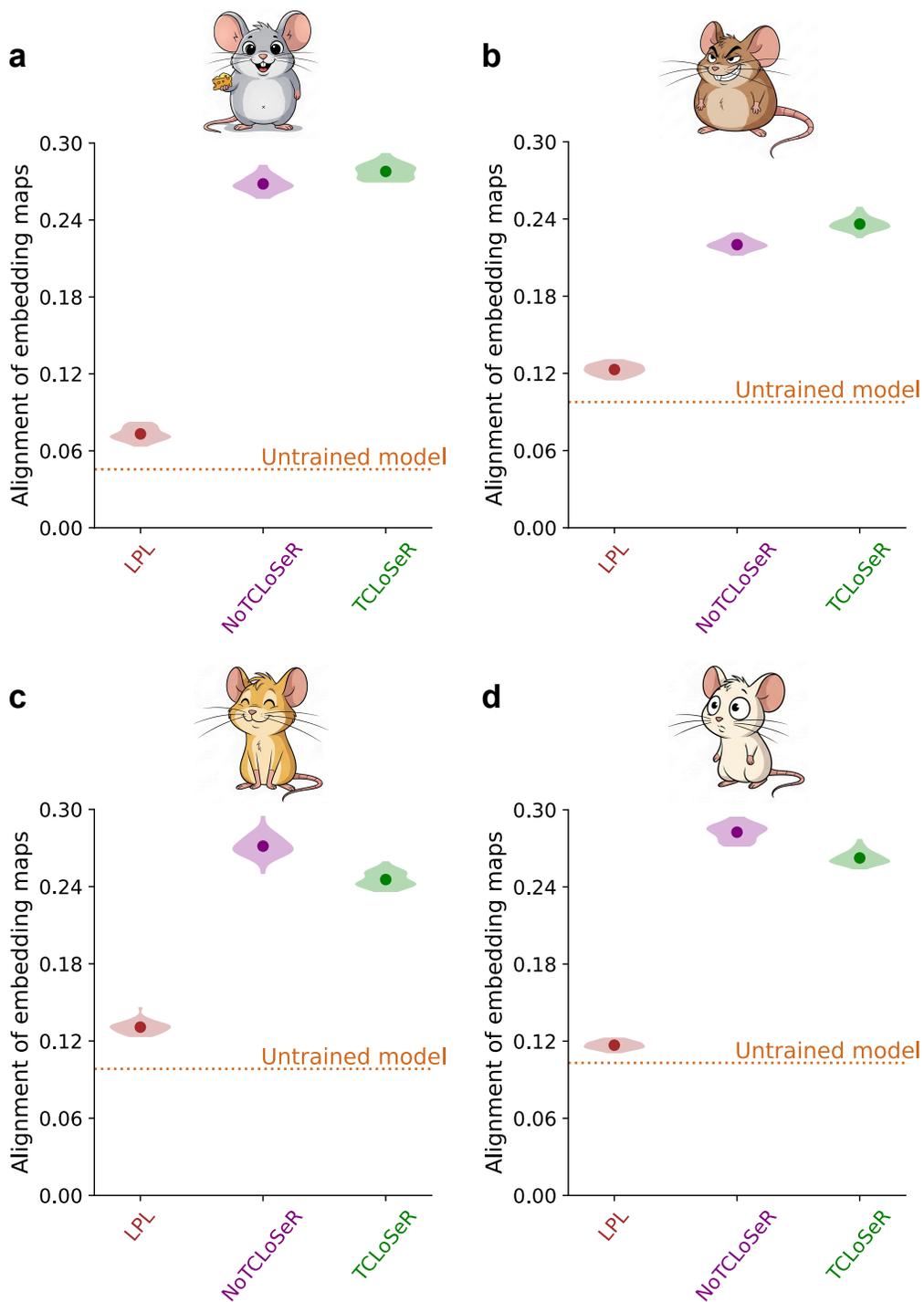}
     \caption{\textbf{Alignment of encoders trained for temporal contiguity and cross-supervision} 
     \textbf{(a)} Violinplot of the average alignment of embedding maps of MLP encoders is shown for different encoder types (see \nameref{Methods: Alignment of embedding maps}). Calculated over 30 seeds.
     \textbf{(b-d)} Same as \textbf{a} for different mice session (ids 755434585, 757216464, 757970808).
     }
     \label{fig_sup_neur_align}
\end{figure*}

\end{document}